\documentclass[12pt]{article}
\usepackage{putex}
\usepackage{graphicx}

\usepackage{amsfonts,amsthm}
\usepackage{float}
\usepackage[english]{babel}
\usepackage[utf8]{inputenc}
\usepackage{slashed}
\usepackage{subcaption} 
\usepackage{booktabs}   
\usepackage{hyperref}
\hypersetup{colorlinks}

\begin{document}
\preprint{PUPT-2486}

\title{Segmented Strings in $AdS_3$}
\authors{Nele Callebaut,$^\PU$ Steven S. Gubser,$^\PU$ Andreas Samberg,$^{\UH,\EM}$ and Chiara Toldo$^\CU$}
\institution{PU}{\hskip-0.05in${}^1$Joseph Henry Laboratories, Princeton University, Princeton, NJ 08544, USA}
\institution{UH}{\hskip-0.05in${}^2$Institut f\"ur Theoretische Physik, Ruprecht-Karls-Universit\"at Heidelberg,\cr Philosophenweg~16, 69120~Heidelberg, Germany}
\institution{EM}{\hskip-0.05in${}^3$ExtreMe Matter Institute EMMI, GSI Helmholtzzentrum 
f\"ur Schwerionenforschung,\cr Planckstra{\ss}e~1, 64291~Darmstadt, Germany}
\institution{CU}{\hskip-0.05in${}^4$Department of Physics, Columbia University, 538 West 120th Street,\cr New York, NY 10027, USA}
\abstract{
We study segmented strings in flat space and in $AdS_3$.  In flat space, these well known classical motions describe strings which at any instant of time are piecewise linear.  In $AdS_3$, the worldsheet is composed of faces each of which is a region bounded by null geodesics in an $AdS_2$ subspace of $AdS_3$.  The time evolution can be described by specifying the null geodesic motion of kinks in the string at which two segments are joined.  The outcome of collisions of kinks on the worldsheet can be worked out essentially using considerations of causality.  We study several examples of closed segmented strings in $AdS_3$ and find an unexpected quasi-periodic behavior.  We also work out a WKB analysis of quantum states of yo-yo strings in $AdS_3$ and find a logarithmic term reminiscent of the logarithmic twist of string states on the leading Regge trajectory.}
\date{August 2015}

\maketitle

\tableofcontents

\section{Introduction}

Strings in flat space are simple because their dynamics is controlled by a free conformal field theory (CFT).  Strings in general curved spacetimes are relatively intractable because, even if the constraint of conformal invariance is obeyed, the worldsheet CFT is too complicated to solve.  Strings in anti-de Sitter (AdS) space are a happy medium, where the worldsheet CFT is interacting but is understood to be integrable.  For entry points into the large literature on integrability, see for example \cite{Beisert:2010jr,Tseytlin:2010jv}.  Classical string solutions in AdS have been treated systematically when their motion is a rigid rotation: As reviewed in \cite{Tseytlin:2010jv}, one can then map the problem of finding their shape into an integrable one-dimensional Neumann model describing some variant of an oscillator on a sphere.  This method, together with some extensions, allows for the construction of quite a variety of string states.  Other methods, including Pohlmeyer reduction, B\"acklund transformations, coset methods, and the dressing method, have been used to produce an extensive and varied collection of classical solutions: See for example \cite{Gubser:2002tv,Arutyunov:2003rg,Kruczenski:2004wg,Kazakov:2004nh,Kalousios:2006xy,Alday:2007hr,Grigoriev:2007bu,Mikhailov:2007xr,Jevicki:2007aa,Alday:2009yn,Alday:2010vh}.

In this paper we focus on a class of classical string motions in $AdS_3$ which can be analyzed by elementary means.  The string configurations can be thought of as segmented, where each segment stretches along the intersection of a time-slice in $AdS_3$ with an $AdS_2$ subspace.  Tracking the motion of the string is as simple as tracking the lightlike motion of the points where the segments join together.  An analogous class of classical string motions in flat space has been studied extensively starting with \cite{Artru:1979ye} and is central to the Lund model of mesons \cite{Andersson:1983ia}.  In the Lund model, a basic ingredient is the yo-yo string, where a string starts at rest stretched between two points, and then contracts inward until its endpoints meet.  Then it expands outward again until it regains its original length, and contracts once again.  An interesting point is that any piecewise linear string can be constructed as a union of boosted yo-yos, joined up so that one endpoint matches with the next.  There is a small caveat: Generically, we may expect that a string solution does not have finite momentum localized at a point, but yo-yos do, except when they reach their full length.  We will adopt a simple treatment where we think of dropping the endpoint momentum when two yo-yo solutions are joined at their endpoints.\footnote{This approach imposes some limits on the class of string solutions we are able to study.  For example, we cannot in this way study a string doubled over on itself, because as the doubled-over part contracts, finite momentum collects at the kink.}  Yo-yo strings were studied in $AdS_5$ \cite{Ficnar:2013wba}.  For earlier related work see also \cite{Klebanov:2006jj}.

A variant of the yo-yo is possible in $AdS_3$, where a string undergoes yo-yo motions while its center of mass stays fixed at the center of $AdS_3$.  Just as we can boost a yo-yo solution in flat space, we can apply a (global) conformal transformation to a yo-yo solution in $AdS_3$.  Even after an arbitrary global conformal transformation, the yo-yo will move in some $AdS_2$ subspace of $AdS_3$.  We can assemble conformally transformed yo-yos in the same way as in flat space, dropping endpoint momentum where yo-yo endpoints join together.  In this way we obtain the classical closed string motions we are interested in.  Treating open string motions similarly requires inclusion of localized momentum and is beyond our current scope.

The discussion of the previous two paragraphs leaves out an important aspect of time evolution.  If we are considering segmented strings with kinks where segments join, then as the string propagates forward in time, we have to ask what happens when kinks collide.  The answer, it turns out, can be worked out almost completely from considerations of causality.  As a warm-up, in section~\ref{FLAT} we will explain how segmented strings work in flat space.  The results are trivial in the sense that we are only recovering the results of free field theory.  The main results are in section~\ref{ADS}, where we explain the evolution of segmented strings in $AdS_3$ as well as a partial account of initial conditions.  Section~\ref{ENERGY} is devoted to energy considerations, including computation of the conserved energy of strings in $AdS_3$ and a WKB analysis of energy levels.  In section~\ref{EXAMPLES} we present some examples of motions of segmented strings in $AdS_3$ which are non-periodic.

\section{Piecewise linear strings}
\label{FLAT}

Classical string trajectories in flat space must take the form
 \eqn{RightLeftExpand}{
  X^\mu(\tau,\sigma) = {1 \over 2} \left( Y^\mu_R(\tau-\sigma) + Y^\mu_L(\tau+\sigma)
    \right) \,.
 }
As written, \eno{RightLeftExpand} is the general solution to the flat space string equations of motion written in conformal gauge.  The Virasoro constraints are satisfied if $Y^\mu_R(\xi)$ and $Y^\mu_L(\xi)$ are null trajectories in spacetime.  Any such trajectory can be approximated by a piecewise linear null trajectory.  If we do this for $Y^\mu_R$ and $Y^\mu_L$, then the $X^\mu(\tau,\sigma)$ are piecewise linear functions of $\tau$ and $\sigma$.  Such string motions are precisely the segmented strings we are interested in.

On any time slice (meaning a slice of constant $X^0$), a segmented string must be a union of line segments, each of whose center of mass follows a timelike trajectory while its endpoints move at the speed of light.  The endpoints are kinks where the direction of spatial extent of the string along a fixed timeslice can change.  Thus we see that segmented strings can indeed be regarded as a union of boosted yo-yo solutions, except without endpoint momentum.  The time evolution of the string can easily be described by tracking the kinks until two kinks collide.  As we will explain, we can determine the ``outcome'' of such a collision, namely the direction of two new kinks that come out of the collision, based on knowing only the geometry of the string worldsheet in a small neighborhood of the collision.  In other words, to know the entire motion of a piecewise linear string, one has only to track the motion and interactions of finitely many piecewise null geodesics on its worldsheet, using only local information at the kinks.

The conclusions of the previous paragraph are trivial in the sense that we are only restructuring the solution \eno{RightLeftExpand}.  But it is an appealing thought that the description of string dynamics that we will give generalizes to the far less trivial problem of strings in curved spacetime.  In short, we make the following conjectures:
 \begin{itemize}
  \item There are classical string trajectories in suitable curved spacetimes which can be completely described by tracking the piecewise null trajectories of finitely many points on the string worldsheet, together with the direction of the string worldsheet running into and out of each of these points.
  \item The space of all the trajectories capable of such description is dense in the space of all possible classical string motions.
 \end{itemize}
By ``suitable curved spacetimes'' we mean spacetimes in which strings can consistently propagate---which in terms of conformal field theory means target spaces whose beta functionals vanish.

Before exploring these conjectures further, let's articulate in flat space exactly how a description of classical string motion works when we specify data only in the vicinity of several lightlike kinks.  Let these kinks be numbered $i=1,2,\ldots,N$, and let their spatial positions be
 \eqn{KinkPositions}{
  \vec{x}_i(t) = \vec{v}_i t + \vec{d}_i \,.
 }
(Of course, the linear form \eno{KinkPositions} of $\vec{x}_i(t)$ only applies piecewise: That is, $\vec{v}_i$ and $\vec{d}_i$ change discontinuously at special times when kinks collide.)  Assume further that the string (which we take to be oriented) runs into kink $i$ along a spatial unit vector $\vec\ell_i$, and out of it along a spatial unit vector $\vec{r}_i$.  We can uniquely decompose
 \eqn{ellDecompose}{
  \vec{v}_i = v_{iL} \vec\ell_i + \vec{v}_{iT} = v_{iR} \vec{r}_i + \vec{v}_{iS}
 }
where $\vec{v}_{iT} \perp \vec\ell_i$ and $\vec{v}_{iS} \perp \vec{r}_i$.  The velocity $\vec{v}_{iT}$ ($\vec{v}_{iS}$) is the transverse velocity of the string running into (out of) the kink, and because this velocity must be timelike, the signed real quantity $v_{iL}$ ($v_{iR}$) must be non-zero.  Similarly, $v_{iR}$ must be non-zero.  We make an important restriction on the states we allow by requiring the product $v_{iL} v_{iR}$ to be positive.  Physically, the kink must either be moving to the right along the string, in which case $v_{iR} > 0$, or left along the string, in which case $v_{iR} < 0$.  If we considered cases where $v_{iL} v_{iR} < 0$, then we have to allow finite momentum at the kink, whose linear increase or decrease with time would be part of the dynamics.

An additional constraint is that the string running out of kink $i$ must connect with the string running into kink $i+1$: That is, $\vec{r}_i = \vec\ell_{i+1}$ must be the unit vector in the direction of $\vec{x}_{i+1}(t) - \vec{x}_i(t)$.  There are now three subcases to consider:
 \begin{itemize}
  \item If kink $i$ is right-moving while kink $i+1$ is left-moving, then the two kinks will collide at some future time $t_*$, and one can see that $\vec{r}_i = \vec\ell_{i+1}$ is the unit vector parallel to $\vec{v}_i - \vec{v}_{i+1}$.
  \item If kink $i$ is left-moving while kink $i+1$ is right-moving, then the two kinks came out of a collision at some earlier time $t_*$, and $\vec{r}_i = \vec\ell_{i+1}$ is the unit vector parallel to $\vec{v}_{i+1} - \vec{v}_i$.
  \item If both kinks are right-moving, or both are left-moving, then $\vec{v}_i = \vec{v}_{i+1}$, and $\vec{r}_i = \vec\ell_{i+1}$ is the unit vector parallel to $\vec{d}_{i+1} - \vec{d}_i$.
 \end{itemize}
Here and below, what we mean by two vectors $\vec{a}$ and $\vec{b}$ being parallel is that $\vec{a} = \lambda \vec{b}$ for some positive real number $\lambda$.  If instead $\lambda$ were negative, we would refer to $\vec{a}$ and $\vec{b}$ as anti-parallel.

Thus far, we have only described the time evolution of kinks in the absence of collisions between kinks, along with consistency conditions that must be satisfied throughout that time evolution.  To describe a collision, first note that we must start with a right-moving kink to the left of a left-moving kink and end up instead with a left-moving kink to the left of a right-moving kink.  The key point in predicting the outcome of a collision of kinks is that the spatial orientation and transverse motion of the string running into the leftmost kink cannot change: That is, $\vec\ell_i$ and $\vec{v}_{iT}$ are unaltered by the collision.  This conclusion can be reached on grounds of causality applied to the string segment running into the leftmost kink.  Whatever happens at the collision, a piece of string at some finite distance away from the collision cannot find out about it until the newly formed kink traverses back across the string at the speed of light.  By a similar argument, $\vec{r}_{i+1}$ and $\vec{v}_{i+1,S}$ are unaltered.  What does happen at the collision is that kinks pass through one another, or bounce off each other, and the leftmost one starts moving back along the worldsheet (i.e.~to the left) while the rightmost one moves forward along the worldsheet (i.e.~to the right).  The only way this can happen is if the kink's longitudinal velocities after the collision are given by
 \eqn{vLSwitch}{
  \tilde{v}_{iL} = -v_{iL} \qquad\qquad \tilde{v}_{i+1,R} = -v_{i+1,R}
 }
By demanding that $\vec{x}_i(t)$ and $\vec{x}_{i+1}(t)$ are continuous at the time $t=t_*$ of the collision, we straightforwardly obtain the relations
 \eqn{dAlter}{
  \tilde{\vec{d}}_i = \vec{d}_i + 2 v_{iL} \vec\ell_i t_* \qquad\qquad
  \tilde{\vec{d}}_{i+1} = \vec{d}_{i+1} + 2 v_{i+1,R} \vec{r}_{i+1} t_* \,.
 }
Finally, we should ask in what spatial direction the string between the kinks runs along after the collision.  The answer is that this direction, $\vec{r}_i = \vec\ell_{i+1}$, is the unit vector parallel to $\tilde{\vec{v}}_{i+1} - \tilde{\vec{v}}_i$.

In summary: Equations \eno{KinkPositions}-\eno{dAlter} are sufficient to determine the motion of piecewise linear strings assuming that there is no localized energy or momentum at the kinks between linear segments.  We could regard the case where there is localized energy or momentum at a kink as a limit of a situation where a very short segment of string propagates at nearly the speed of light.

\section{Strings in $AdS_3$}
\label{ADS}

Let us now consider strings in $AdS_3$, which we describe as the locus of points in ${\bf R}^{2,2}$ satisfying the equation
 \eqn{AdSembedded}{
  -(Y^{-1})^2 - (Y^0)^2 + (Y^1)^2 + (Y^2)^2 = -1 \,.
 }
We may parametrize the locus with coordinates $(\tau,\rho,\phi)$ as
 \eqn{PolarCoords}{
  \begin{pmatrix} Y^{-1} \\ Y^0 \\ Y^1 \\ Y^2 \end{pmatrix} = 
    \begin{pmatrix} \cosh\rho \cos\tau \\ \cosh\rho \sin\tau \\
      \sinh\rho \cos\phi \\ \sinh\rho \sin\phi \end{pmatrix} \,.
 }
$AdS_3$ is in fact the universal covering space of the hyperboloid \eno{AdSembedded}; in practical terms this means that as global time $\tau$ evolves forward by $2\pi$, we do not return to the same point, but instead to a new sheet of the covering space.

\subsection{The yo-yo in $AdS_3$}

As in the introduction, we will start our discussion of string motions in $AdS_3$ with the yo-yo.  Assume that at time $\tau=0$, the string starts at $\rho=0$, and after some time it expands to stretch from $(\rho,\phi) = (\rho_*,\pi)$ to $(\rho,\phi) = (\rho_*,0)$.  This string lies wholly in the $AdS_2$ submanifold specified by intersecting the hyperboloid \eno{AdSembedded} with the plane $Y_2=0$.  We rewrite the equation $Y_2=0$ as
 \eqn{AdStwo}{
  k_A Y^A = 0
 }
where $k^A = (0,0,0,1)$ and we raise indices with the natural metric $\diag\{-1,-1,1,1\}$ on ${\bf R}^{2,2}$.  The right-moving endpoint of the yo-yo travels along the null geodesic
 \eqn{TwoGeodesics}{
  Y^A(\xi) = h^A + v^A \xi
 }
where $h^A = (1,0,0,0)$ and $v^A = (0,1/\sqrt{2},1/\sqrt{2},0)$.  (An important observation is that \eno{TwoGeodesics} is a null geodesic both in the embedding spacetime ${\bf R}^{2,2}$ and on the hyperboloid $AdS_3$ defined by \eno{AdSembedded}.)  Observe that $h$, $k$, and $v$ are mutually orthogonal, and if we include also $u^A = (0,1/\sqrt{2},-1/\sqrt{2},0)$ we have a basis
 \eqn{BBasis}{
  B = (h,k,u,v)
 }
which satisfies the relations
 \eqn[c]{BSatisfies}{
  h^2 = -1 \qquad k^2 = 1 \qquad u^2 = 0 \qquad v^2 = 0  \cr
  h \cdot k = h \cdot u = h \cdot v = 0 \qquad
  k \cdot u = k \cdot v = 0 \qquad
  u \cdot v = -1 \,.
 }
An overall multiplicative factor on $v$ can be adjusted at will, provided we correspondingly rescale $u$ to maintain the relation $u \cdot v = -1$.  Define now $t$ as the unit vector field on $AdS_3$ in the direction of $d/d\tau$.  We require that $t \cdot v$ and $t \cdot u$ are negative; that is, $v$ and $u$ are future-directed.  If it is desired to fix the freedom of adjusting $v$ by an overall multiplicative factor, we may additionally demand $t \cdot v = -1/\sqrt{2}$.  Note that in general it is not consistent to demand additionally $t \cdot u = -1/\sqrt{2}$.

Observe for the basis indicated below \eno{TwoGeodesics} that 
 \eqn{RightHanded}{
  v^A = -\epsilon^A{}_{BCD} h^B k^C v^D
 }
where $\epsilon_{ABCD}$ is antisymmetric with $\epsilon_{-1,0,1,2} = 1$.  If instead we started with $v^A = (0,1/\sqrt{2},$ $-1/\sqrt{2},0)$, with $k^A = (0,0,0,1)$ and $h^A = (1,0,0,0)$ as before, then to satisfy \eno{BSatisfies} we would need $u^A = (0,1/\sqrt{2},1/\sqrt{2},0)$, which results in a coordinate system with the opposite orientation: That is, the explicit minus sign would be absent the right hand side of \eno{RightHanded}.  We will refer to a basis $B$ satisfying \eno{RightHanded} as having orientation $\sigma(B) = +1$, while if the explicit minus were absent, we would say $\sigma(B) = -1$.

In summary, the right-moving kink at the tip of yo-yo can be characterized by a basis $B_L$ with orientation $+1$, while the left-moving kink can be characterized by a basis $B_R$ of orientation $-1$; and the bases $B_L$ and $B_R$ share the same vectors $h$ and $k$, while the forward-directed null vectors $u$ and $v$ are flipped between $B_L$ and $B_R$, up to possible rescalings of $v$ (and hence $u$) which correspond to rescalings of $\xi_L$ and $\xi_R$.  The choice of the spacelike vector $k$ specifies the $AdS_2$ subspace in which the yo-yo propagates; the timelike vector $h$ specifies the moment at which the two ends of the yo-yo begin to separate; and of course the vectors $v_L$ and $v_R$ indicate the null directions within $AdS_2$ in which the right-moving and left-moving kinks propagate.

It is easy enough to predict how the yo-yo will evolve because it remains always in the same $AdS_2$ subspace.  Namely, the endpoints start off with some definite momentum, which for simplicity we assume to be equal and opposite.  (More technically, the sum of the energy momentum vectors of the two endpoints at the initial point $h$ where they coincide is in the direction of $t$; if this is not the case we can change $t$ by a conformal transformation to make it so, amounting to some $SO(2,2)$ transformation on ${\bf R}^{2,2}$.)  The energy and momentum bleed off from the endpoint into the bulk of string as it extends in the spatial direction of $AdS_2$ until no energy remains; then the endpoints snap back and propagate along null trajectories back to a position in $AdS_2$ which is $h$ plus some positive multiple of $t$.

It will pay to give a slightly more explicit description of this motion in terms of the bases $B_L$ and $B_R$.  For either basis (dropping the subscript for notational simplicity), the evolution with affine parameter $\xi$ along a null trajectory is given by
 \eqn{AdvanceBasis}{
  h(\xi) = h + \xi v \qquad k(\xi) = k \qquad
  u(\xi) = u - \xi h - {\xi^2 \over 2} v \qquad
  v(\xi) = v \,,
 }
where $h$, $k$, $u$, and $v$ on the right hand sides are understood to be the initial values at $\xi=0$.  The evolution $h(\xi)$ is what we mean by advancing along a null trajectory.  Keeping $k$ and $v$ constant simply means that we stay on a definite null trajectory within a definite $AdS_2$ subspace.  The form of $u(\xi)$ is forced upon us by the requirement of preserving the orthogonality relations \eno{BSatisfies}.  Alternatively, we may arrive at this evolution by parallel transporting $u$ along the null trajectory, using the natural connection on $AdS_3$ (not the trivial connection on ${\bf R}^{2,2}$).  Note that the condition $t \cdot v = -1/\sqrt{2}$ is not preserved by the evolution \eno{AdvanceBasis}: A direct calculation of $t$ along the null geodesic yields
 \eqn{tAdvance}{
  t(\xi) = {u + v - \xi h \over \sqrt{2+\xi^2}} \,,
 }
so $t(\xi) \cdot v(\xi) = -1/\sqrt{2+\xi^2}$.

Snap-back occurs at some value $\xi = \xi_*$ which depends on the initial energy and momentum carried by the endpoints.  Snap-back is easy to describe in terms of basis vectors: one simply swaps $v$ and $u$, keeping $h$ and $k$ the same.  As a result, the orientation of each basis flips.  We may then reapply the time evolution \eno{AdvanceBasis} to each new basis, with the result that the endpoints travel back toward one another as summarized previously.  When they meet, they should be understood to pass through one another, separating once more to the same maximum distance before snapping back again.

\subsection{Closed strings in $AdS_3$}

The generalization of the yo-yo that we propose is the closest analog to piecewise linear strings that can be achieved in $AdS_3$.  Namely, on a particular slice of constant global time $\tau$, let there be $N$ kinks, which are cyclically arranged in the order in which the string passes through them.  The string is, by assumption, oriented and closed, so we think of the string as starting at kink $i=1$, proceeding to $i=2$, and so on up to kink $i=N$, and finally back to kink $i=1$.  For simplicity, we stipulate as before that there is no localized energy or momentum at the kinks.  Each kink is ``decorated'' with an ``extended basis,'' call it $B_i$, which comprises {\it six} vectors in ${\bf R}^{2,2}$: dropping the index $i$ for simplicity,
 \eqn{BiForm}{
  B = (h,j,k,w,u,v) \,.
 }
In addition to the relations \eno{BSatisfies}, we demand also
 \eqn[c]{BiExtras}{
  j^2 = 1 \qquad w^2 = 0  \cr
  h \cdot j = h \cdot w = 0 \qquad j \cdot w = j \cdot v = 0 \qquad w \cdot v = -1 \,.
 }
In short, $(h,k,u,v)$ is a basis in the sense of \eno{BSatisfies}, characterizing the string running out from the kink, and $(h,j,w,v)$ is another such basis characterizing the string running into the kink.  We also require that these two bases have the same orientation in the sense explained following \eno{RightHanded} (with $w$ future-directed like $v$ and $u$ are); we will refer to this orientation as $\sigma(B) = \pm 1$.  Explicitly, if \eno{RightHanded} holds as written, then we must also have
 \eqn{RightHandedAgain}{
  v^A = -\epsilon^A{}_{BCD} h^B j^C v^D \,.
 }
If $\sigma(B) = +1$, then the corresponding kink is right-moving in the sense that the string segment running into it (i.e.~from the left in worldsheet terms) is lengthening with increasing global time $\tau$, while the string segment running out of it (i.e.~to the right in worldsheet terms) is shortening.  If $\sigma(B) = -1$, then the kink is left-moving on the worldsheet.

Time evolution of the string worldsheet is a generalization of the discussion of the yo-yo.  The formula \eno{AdvanceBasis} may be augmented by the rules
 \eqn{jwAdvance}{
  j(\xi) = j \qquad w(\xi) = w - \xi h - {\xi^2 \over 2} w
 }
and applied to each basis.  We must arrange initial conditions so that adjacent kinks with opposite orientation collide either in the future or in the past, and so that the kinks' trajectories lie on a common $AdS_2$ subspace.  Let's treat the case of a right-moving kink $1$ and a left-moving kink $2$; then the collision must happen in the future.  We first use \eno{AdvanceBasis} and \eno{jwAdvance} to advance the kinks to the collision point.  For simplicity we now drop all reference to $\xi$ and simply assume that the extended bases $B_1 = (h_1,j_1,k_1,w_1,u_1,v_1)$ and $B_2 = (h_2,j_2,k_2,w_2,u_2,v_2)$ satisfy
 \eqn{CollisionConditions}{
  h_1 = h_2 \qquad k_1 = j_2 \,.
 }
To work through the logic of a collision, it helps to consider first the ``outer'' segments of string, namely the string segment running into kink $1$ and the segment running out of kink $2$.  These segments must remain in the same spatial orientation---that is, $j_1$ and $k_2$ are unchanged.  The reason is that if we go out a little way along either of the outer segment, then we are spacelike separated from the collision itself, and nothing about the collision can affect the motion of the string where we are.  Of course, $h_1$ and $h_2$ are also unchanged.
Let's improve notation and write
 \eqn{jkTilde}{
  \tilde{h}_1 = \tilde{h}_2 = h_1 = h_2 \qquad 
    \tilde{j}_1 = j_1 \qquad \tilde{k}_2 = k_2 \,,
 }
where a tilde is used to indicate data relating to after the collision.  We may further reason that
 \eqn[c]{vTilde}{
  \tilde{v}_1 = w_1 \qquad \tilde{w}_1 = v_1 \qquad
  \tilde{v}_2 = u_2 \qquad \tilde{u}_2 = v_2 \,.
 }
The justification for $\tilde{v}_1 = w_1$ is that after the collision, it must travel in a forward-directed null direction within the $AdS_2$ subspace orthogonal to $j_1$, and the only such direction other than $v_1$ is $w_1$.  Then we must have $\tilde{w}_1 = v_1$ to maintain orthogonality relations.  The third and fourth equations of \eno{vTilde} can be justified similarly.  This reasoning is perfectly analogous to the description of snap-back for the yo-yo.  Note that due to swapping $v_1$ and $w_1$, now kink $1$ has orientation $-1$ (that is, it is left-moving), while kink $2$ has flipped its orientation to $+1$, i.e. right-moving.  Thus we preserve the order of the kinks through the collision; intuitively, we think of the kinks as bouncing back off one another rather than passing through one another.

In order to figure out what happens to the ``inner'' segment of string after the collision, let's first note that we know $\tilde{v}_1$ and $\tilde{v}_2$ from \eno{vTilde}.  These two null vectors at the collision point uniquely determine the $AdS_2$ subspace in which the inner segment of string must lie.  Because $\tilde{v}_1$ and $\tilde{v}_2$ form a null basis for the tangent space of this $AdS_2$ subspace at the collision point, it must be that $\tilde{u}_1$ is some multiple of $\tilde{v}_2$ and $\tilde{w}_2$ is some multiple of $\tilde{v}_1$; only then can $(\tilde{v}_1,\tilde{u}_1)$ and $(\tilde{v}_2,\tilde{w}_2)$ also be bases for the same $AdS_2$ subspace at the collision point.  More specifically:
 \eqn{wuTilde}{
  \tilde{u}_1 = -{\tilde{v}_2 \over \tilde{v}_1 \cdot \tilde{v}_2} \qquad
  \tilde{w}_2 = -{\tilde{v}_1 \over \tilde{v}_1 \cdot \tilde{v}_2} \,,
 }
where the denominators enforce the relevant orthogonality relations.  At this point, the only vectors left are $\tilde{k}_1$ and $\tilde{j}_2$.  These vectors determine the $AdS_2$ subspace in which the inner string propagates after the collision.  But we already know which $AdS_2$ we want: it is the one through the collision point whose tangent space has basis $(\tilde{v}_1,\tilde{u}_1)$, or equivalently $(\tilde{v}_2,\tilde{w}_2)$.  It is straightforward to check that
 \eqn{jkInner}{
  \tilde{k}_1^A = \tilde{j}_2^A = 
    -\epsilon^A{}_{BCD} \tilde{h}_1^B \tilde{u}_1^C \tilde{v}_1^D
 }
is the unique choice that will satisfy the orthogonality relations as well as \eno{RightHanded} and \eno{RightHandedAgain} for kink 2 (which is right-moving), and the same conditions without the explicit minus signs for kink 1 (which is left-moving).

In \eno{CollisionConditions} we stated the minimal set of preconditions on the bases $B_1$ and $B_2$ before the collision required in order for the discussion \eno{jkTilde}-\eno{jkInner} to make sense.  In fact, if we run the logic of the collision in reverse, we can deduce some additional preconditions relating to the inner segment of string before the collision:
 \eqn{ExtraConditions}{
  u_1 = -{v_2 \over v_1 \cdot v_2} \qquad w_2 = -{v_1 \over v_1 \cdot v_2} \qquad
  k_1^A = j_2^A = -\epsilon^A{}_{BCD} h_1^B u_1^C v_1^D \,.
 }

\subsection{Initial conditions}\label{icSubsection}

We have laid out string evolution so far without specifying exactly what initial conditions one is supposed to evolve from, say on the time-slice $\tau=0$.  A quick way out is to allow precisely those initial conditions which, if evolved forward in time, lead to collisions where the preconditions \eno{CollisionConditions} and \eno{ExtraConditions} are satisfied for every pair of colliding vertices, and if evolved backward in time, lead to collisions where for any pair of colliding vertices, call them kinks $1$ and $2$, we have $\tilde{h}_1 = \tilde{h}_2$, $\tilde{k}_1 = \tilde{j}_2$, and \eno{wuTilde}.\footnote{This set of constraints is sufficient provided we alternate right-moving and left-moving kinks.  If there are, for example, several left-moving kinks in a row, then we have to put constraints going forward in time on the leftmost one, which undergoes a collision with the right-moving kink just to its left, and then constrain the next-to-leftmost left-moving kink in terms of the right-moving kink emerging from said collision.}  In practice, we would like a more constructive account of allowed initial conditions.

We do not have complete results, but we will offer here a construction of initial conditions based on the assumption that one single collision of kinks occurs at time $\tau = 0$.  Let there be $N$ kinks total, with $N$ stipulated to be an even number, and let the two kinks undergoing a collision at $\tau=0$ be kinks number $N-1$ and $N$.  Specify first the desired positions $h_1,h_2,\ldots,h_{N-1}$, $h_N$ at time $\tau=0$, with $h_{N-1} = h_N$ by assumption.  Note that each position may be expressed as
 \eqn{hForm}{
  h_i = (h_i^{-1},0,\vec{h}_i) \qquad\hbox{where}\qquad
   h_i^{-1} = \sqrt{1 + \vec{h}_i^2} \,,
 }
and we use the short-hand $\vec{X}$ to mean the spatial components $(X^1,X^2)$ of any vector in ${\bf R}^{2,2}$.  The positive sign in front of the square root in \eno{hForm} must be chosen because that is what corresponds to $\tau = 0$ (rather than $\tau = \pi$).  Thus in \eno{hForm}, the quantities that can be freely specified are $\vec{h}_1,\vec{h}_2,\ldots,\vec{h}_{N-1}$, for a total of $2(N-1)$ real free parameters, subject to the condition that adjacent kinks must be distinct---and that includes the constraint that $\vec{h}_{N-1}=\vec{h}_N$ differs from $\vec{h}_1$.

Next, specify the spatial components $\vec{v}_1$ of the velocity $v_1$.  The condition $h_1 \cdot v_1 = 0$ allows us to compute
 \eqn{vmOneOne}{
  v_1^{-1} = {\vec{h}_1 \cdot \vec{v}_1 \over h_1^{-1}} \,,
 }
where we used $\tau=0$ to conclude $h_0=0$.  The condition $v_1 \cdot v_1 = 0$ is now a quadratic equation for $v_1^0$ whose solution is
 \eqn{vZeroOne}{
  v_1^0 = \sqrt{\vec{v}_1^{\,2} - (v_1^{-1})^2} \,.
 }
The quantity inside the square root must be positive because of the Schwarz inequality applied to \eno{vmOneOne} together with $h_1^{-1} = \sqrt{1 + \vec{h}_1^2}$.  We must choose the positive sign on the square root in \eno{vZeroOne} because we want $v_1$ to be future directed.  Note that no consideration in this paragraphs restricts $\vec{v}_1$ in any way, except that it should be non-zero in order for $v_1$ as a whole to be non-zero.  Thus $\vec{v}_1$ adds two more free real parameters to the initial conditions, for a total of $2N$.  We will see as we go on with our construction that not all values of $\vec{v}_1$ are allowed.

The plan now is to figure out what $v_2$ must be, then $v_3$, and so forth up to $v_{N-1}$.  To determine $v_2$, we impose the condition that kinks $1$ and $2$ must collide at some future time $\tau_1$, with no other kinks colliding with either $1$ or $2$ in the interval $0 < \tau < \tau_1$.  (The case where kinks $1$ and $2$ collide at some time in the past proceeds almost identically.)  Then we must be able to solve the equations
 \eqn{OneTwoCollide}{
  h_1 + \xi_1 v_1 = h_2 + \zeta_2 v_2 \qquad v_2 \cdot v_2 = h_2 \cdot v_2 = 0 \,.
 }
Moreover, the affine times $\xi_1$ and $\zeta_2$ that elapse before kinks collide must be positive.  We are free to rescale $v_2$ by a positive factor; let us use this freedom to set $\zeta_2 = \xi_1$.  Defining
 \eqn{DeltaH}{
  \Delta h_i = h_{i+1} - h_i \qquad\hbox{for}\qquad 1 \leq i < N \,,
 }
one can easily check that the unique solution for $\xi_1=\zeta_2$ and $v_2$ is
 \eqn{vTwoForm}{
  \xi_1=\zeta_2 = {(\Delta h_1)^2 \over 2 \Delta h_1 \cdot v_1} \qquad
    v_2 = v_1 - 2 {\Delta h_1 \cdot v_1 \over (\Delta h_1)^2} \Delta h_1 \,.
 }
Note that our assumption that $h_1$ and $h_2$ are distinct implies $(\Delta h_1)^2 > 0$, since this inequality is the statement that $h_1$ and $h_2$ are spacelike separated.  So we must have $\Delta h_1 \cdot v_1 > 0$, which can be re-expressed as the constraint
 \eqn{vOnePositivity}{
  h_2 \cdot v_1 > 0 \,,
 }
or, after the use of \eno{vmOneOne},
 \eqn{vOneCondition}{
  \vec{v}_1 \cdot \left( {\vec{h}_2 \over h_2^{-1}} - {\vec{h}_1 \over h_1^{-1}} \right)
    > 0 \,.
 }
We next require that kinks $2$ and $3$ must have collided at some time $\tau_2$ in the past, with no collisions of any other kinks with $2$ or $3$ for $\tau_2 < \tau < 0$, and we use similar manipulations to compute $v_3$.  Likewise we require that kinks $3$ and $4$ will collide at some time in the future (with the usual restriction against collisions with other kinks) and obtain $v_4$---and so forth, with alternating past and future collisions, until we reach kink $N-1$.  The result of all these computations can be summarized by the relations
 \eqn{IterateVelocities}{
  v_{i+1} = v_i - 2 {\Delta h_i \cdot v_i \over (\Delta h_i)^2} \Delta h_i
 }
and
 \eqn{VelocityInequalities}{
  (-1)^{i+1} h_{i+1} \cdot v_i > 0 \qquad 
  (-1)^{i+1} \vec{v}_i \cdot \left( 
    {\vec{h}_{i+1} \over h_{i+1}^{-1}} - {\vec{h}_i \over h_i^{-1}} \right) > 0 \,,
 }
all for $1 \leq i < N-1$.  For fixed $i$, the two inequalities in \eno{VelocityInequalities} are equivalent.

The only kink velocity yet to be specified is $v_N$.  We start as before with the requirement that kinks $N$ and $1$ must have collided at some time $\tau_N$ in the past without having collided with other kinks in the time interval $\tau_N < \tau < 0$:
 \eqn{hvAgain}{
  h_N + \xi_N v_N = h_1 + \zeta_1 v_1 \qquad v_N \cdot v_N = h_N \cdot v_N = 0 \,,
 }
with $\xi_N = \zeta_1 < 0$.  We immediately obtain
 \eqn{vNForm}{
  v_N = v_1 + 2 {\Delta h_N \cdot v_1 \over (\Delta h_N)^2} \Delta h_N
 }
and the equivalent constraints
 \eqn{hvNInequality}{
  h_N \cdot v_1 < 0 \qquad
  \vec{v}_1 \cdot \left( {\vec{h}_N \over h_N^{-1}} - {\vec{h}_1 \over h_1^{-1}}
    \right) < 0 \,,
 }
where we have defined
 \eqn{hNDef}{
  \Delta h_N = h_1 - h_N \,.
 }
The utility of working on a time-slice such that $h_{N-1} = h_N$ is that the equation
 \eqn{hvLast}{
  h_{N-1} + \xi_{N-1} v_{N-1} = h_N + \zeta_N v_N
 }
imposes no further conditions on the $v_i$: this is because $\xi_{N-1} = \zeta_N = 0$.  If instead we required all adjacent $h_i$ to be distinct (including $h_N$ and $h_1$), then we would have to require \eno{IterateVelocities} and \eno{VelocityInequalities} for $i=1,2,\ldots,N$, understanding that $i=N+1$ is identified with $i=1$.  The trouble with this is that the requirement that $v_{N+1}$ as computed from iterating \eno{IterateVelocities} once around the string should match $v_1$ becomes a consistency constraint on $h_1,h_2,\ldots,h_N$ together with $v_1$ which is difficult to solve explicitly, at least for general $N$.  By way of contrast, our approach allows us to freely specify $\vec{h}_1,\vec{h}_2,\ldots,\vec{h}_{N-1}$, and $\vec{v}_1$ and then calculate the remaining $\vec{v}_i$ using the equations \eno{IterateVelocities} and \eno{vNForm}, which are linear in the $\vec{v}_i$.\footnote{In light of the non-linear relation \eno{vZeroOne} between $v_i$ and $\vec{v}_i$, one may question whether \eno{IterateVelocities} and \eno{vNForm} really are linear in the $\vec{v}_i$.  The answer is that they are, because the zero component of $v_i$ doesn't participate in \eno{IterateVelocities} and \eno{vNForm}.}

However, the choice of $\vec{v}_1$ is not really free: We have the inequalities \eno{VelocityInequalities} and \eno{hvNInequality}, and since all the $\vec{v}_i$ are linear functions of $\vec{v}_1$, these inequalities can be cast in the form $\vec{b}_i \cdot \vec{v}_1 > 0$ for some collection of $N-1$ vectors $\vec{b}_i$, where $i=1,2,\ldots,N-2,N$ according to the value of $i$ in the original inequality from which $\vec{b}_i$ arose.  Each of these inequalities restricts $\vec{v}_1$ to a different half-plane in ${\bf R}^2$, so the combination of all of them restricts $\vec{v}_1$ to a wedge bounded by rays starting at the origin.  Depending on the choice of $\vec{h}_1,\vec{h}_2,\ldots,\vec{h}_{N-1}$, this wedge may be non-empty, empty, or in a non-generic case, composed of only a single ray starting at the origin.  Excluding this non-generic case from consideration, we see that the parameter space of initial conditions that we have developed is indeed $2N$-dimensional.

With the construction just described, kinks with odd $i$ are right-moving, while kinks with even $i$ are left-moving.  This applies to kinks $N-1$ and $N$ provided we think of them as describing the state just before their collision at $\tau=0$.  Intuitively, the orientations are as we described because, for example, kink $1$ starts off to the left of kink $2$ and then collides with it---so $1$ must be right-moving while $2$ is left-moving.  To complete the construction of initial conditions, we should make sure that we can construct bases $B_i$ for each kink with the appropriate orientation.  Just before the collision of kinks $1$ and $2$, we may use the formulas \eno{ExtraConditions} to extract $k_1$ and $j_2$.  Because neither $1$ nor $2$ experiences any collisions for $0 < \tau < \tau_1$, $k_1$ and $j_2$ remain unchanged over this interval, and may therefore be used at $\tau=0$.  \eno{ExtraConditions} also tells us $u_1$ and $w_2$ at $\tau = \tau_1$, and these values may be propagated backward using a formula similar to \eno{jwAdvance} to $\tau=0$.  The question of orientations can be settled immediately: given the expressions for $k_1$ and $j_2$ in \eno{ExtraConditions}, it is immediate that \eno{RightHanded} holds for kink $1$ as written, and \eno{RightHandedAgain} holds for kink $2$ without the explicit minus sign.  We may then proceed to the collision between kinks $2$ and $3$ at time $\tau_2 < 0$ and use \eno{jkInner} with $1 \to 2$ and $2 \to 3$ to deduce $k_2$ and $j_3$ (we have dropped tildes since it is understood that we are interested in basis vectors just after the collision).  \eno{wuTilde} with $1 \to 2$ and $2 \to 3$ (and dropping tildes as before) enables us to compute $u_2$ and $w_3$.  Orientation is straightforwardly verified starting with \eno{jkInner}.  The construction of the remaining $B_i$ proceeds similarly, with the collision of $N-1$ and $N$ being simplest of all since one uses \eno{ExtraConditions} directly at $\tau=0$ without having to perform any subsequent evolution of $u_{N-1}$ or $w_N$.

It is worth noting that the alternating orientations we set up in the initial conditions do not generically persist as the string evolves forward in time.  Indeed, the orientations of kinks $N-3,N-2,N-1,N,1,2$ go from perfect alternation (right-left-right-left-right-left) for $\tau$ small and negative to a different pattern (right-left-left-right-right-left) for $\tau$ small and positive, with perfect alternation elsewhere along the string for small enough $\tau$.

\section{Energy considerations}
\label{ENERGY}

As a test of our analysis, we should be able to check that the total energy of a segmented string is conserved over time.  To this end we first consider the action
 \eqn{ActionNoEndpoint}{
  S = -{1 \over 4\pi\alpha'} \int_M d^2 \sigma \, \sqrt{-h} h^{ab}
    \partial_a X^\mu \partial_b X^\nu G_{\mu\nu}
 }
and formulate the worldsheet currents of spacetime energy-momentum:
 \eqn{Pamu}{
  P^a_\mu = -{1 \over 2\pi\alpha'} \sqrt{-h} h^{ab} G_{\mu\nu} \partial_b X^\nu
 }
Then the equation of motion following from \eno{ActionNoEndpoint} is
 \eqn{StringEOM}{
  \partial_a P^a_\mu - \Gamma^\kappa_{\mu\lambda} \partial_a X^\lambda P_\kappa^a = 0 \,,
 }
and if $\zeta_\mu$ is a forward-directed, timelike Killing vector, we can define
 \eqn{Ezeta}{
  E_\zeta = -\int d\sigma \, \zeta^\mu P_\mu^\tau \,,
 }
which is constant, in the sense $\partial_\tau E_\zeta = 0$ when the equations of motion are obeyed.  (We can be more general: For example, if $\zeta_\mu$ is spacelike, then $E_\zeta$ would be a conserved momentum.  Making $\zeta_\mu$ forward-directed and timelike is the case we are interested in currently because we want $E_\zeta$ to be a measure of energy which is positive when $\tau$ increases as one moves forward in spacetime time.)

\subsection{Energy of strings in $AdS_3$}

In $AdS_3$, we can choose $\zeta = \partial_\tau$, and also we identify the worldsheet coordinate $\tau$ with the $AdS_3$ coordinate $\tau$.  Then we arrive at the definition of energy we will use:
 \eqn{Eused}{
  E = -\int d\sigma \, P_\tau^\tau \,.
 }
Next we want to evaluate the integral \eno{Eused} on a segment of string across an $AdS_2$ at a fixed time $\tau$.

As we have seen, an $AdS_2$ face is specified by a vector $k$ with $k \cdot k = 1$.  Suppose we write
 \eqn{kForm}{
  \begin{pmatrix} k^{-1} \\ k^0 \\ k^1 \\ k^2 \end{pmatrix} = 
    \begin{pmatrix} \sinh\kappa \cos\mu \\ \sinh\kappa \sin\mu \\
      \cosh\kappa \cos\theta \\ \cosh\kappa \sin\theta \end{pmatrix} \,.
 }
Then, using \eqref{PolarCoords}, the $AdS_2$ face is all points $Y$ such that
 \eqn{kYForm}{
  k \cdot Y = -\cosh\rho \sinh\kappa \cos(\mu-\tau) + 
    \sinh\rho \cosh\kappa \cos(\theta - \phi) = 0 \,.
 }
Let us first treat the generic case where neither term in the middle expression in \eno{kYForm} vanishes separately.  Then we may solve for $\rho$ in terms of $\phi$:
 \eqn{SolveForRho}{
  \rho = \arccoth\left( {\cos(\theta-\phi) \over \cos(\mu-\tau)} \coth\kappa \right) \,.
 }
Note that the right hand side is a one-to-one function of $\phi$ when the image is required to be real.  As a result, $\phi$ is a good coordinate on the worldsheet segment under consideration.  One can show that
 \eqn{PttComplicated}{
  P_\tau^\tau = {1 \over 2\pi\alpha'} 
   {\cos(\mu-\tau) \sinh\kappa \cos(\theta-\phi) 
     \left[ 1-\cos^2(\mu-\tau) \tanh^2\kappa \right] \over 
     \left[ \cos^2(\theta-\phi) - \cos^2(\mu-\tau) \tanh^2\kappa \right]^{3/2}} \,,
 }
and one must choose the sign on the square root in the denominator to make $P_\tau^\tau$ positive.  The energy integral between two points $h_1$ and $h_2$ (both on the same timeslice as before) is
 \eqn{EComplicated}{
  E_{12} = \left| \int_{\phi_1}^{\phi_2} d\phi \, P_\tau^\tau
    \right| \,.
 }
This integral can be done explicitly in terms of elementary functions, but we do not have a sufficiently simplified expression for the answer to make it useful to record explicitly here.

If both terms in the middle expression in \eno{kYForm} vanish separately, then the treatment becomes a bit more subtle.  In general, $\rho$ will be non-constant along the segment we are interested in, and therefore we must have $\cos(\theta-\phi) = 0$.  This means that $\phi$ is constant, at least on the spatial slice where we are trying to evaluate the energy.  Therefore, $\phi$ is not a good worldsheet coordinate.  It turns out that the best choice of coordinates comes from first rotating space so that $\theta = \pi/2$, so that the string runs along the $\phi=0$ direction, and then introducing new coordinates
 \eqn{SpecialCoords}{
  X^\mu = (\tau,Y^1,Y^2) \,,
 }
where in terms of old coordinates, $(Y^1,Y^2) = (\cosh\rho \cos\phi, \cosh\rho \sin\phi)$.  As usual we identify worldsheet $\tau$ with the $AdS_3$ coordinate $\tau$.  Because the string runs along the $\phi=0$ direction on the timeslice of interest, we can parametrize the spatial direction of the string with $Y^1$.  In short, $\sigma^a = (\tau,Y^1)$.  Noting that $k^A = (\sinh\kappa \cos\mu, \sinh\kappa \sin\mu, 0, \cosh\kappa)$ in the new coordinate system, we find
 \eqn{PttSpecial}{
  P_\tau^\tau = {1 \over 2\pi\alpha'} {1 - (Y^1)^2 \cos^2 (\mu-\tau) \tanh^2 \kappa \over 
    \sqrt{1 - {1 \over 2} \left[ 1 + (Y^1)^2 - (1 - (Y^1)^2) \cos(2(\mu-\tau)) \right]
      \tanh^2\kappa}} \,,
 }
where the square root is chosen to make $P_\tau^\tau$ positive.  The energy may be evaluated as
 \eqn{ESpecial}{
  E_{12} = \left| \int_{h_1^1}^{h_2^1} dY^1 \, P_\tau^\tau \right| \,,
 }
where it is understood that $h_1^1$ and $h_2^1$ are the $Y^1$ components of the endpoints of the string segment on the timeslice of interest.  The indefinite integral of $P_\tau^\tau$ with respect to $Y^1$ can be performed, but again its explicit form is unenlightening.

The total energy of a closed string made out of $N$ segments is
 \eqn{Etot}{
  E_{\rm tot} = \sum_{i=1}^N E_{i,i+1} \,,
 }
where we identify $N+1$ with $1$.  The way we have organized our presentation here, it is not transparent that $E_{\rm tot}$ must be constant; but conservation still follows by a general integration by parts argument on the worldsheet as a whole.

\subsection{An example in $AdS_3$}\label{exampleSquare}

Let's consider an example where there are $N=4$ vertices in $AdS_3$.  Let the initial configuration at $\tau=0$ be a perfect square with corners at $\phi = \pi/2$, $\pi$, $3\pi/2$, and $0$, corresponding to $i=1$, $2$, $3$, and $4$, all at the same value $\rho=\rho_0$.  Thus
 \eqn{hSquare}{
  h_j = \begin{pmatrix} \cosh\rho_0 \\ 0 \\ \sinh\rho_0 \cos {j\pi \over 2} \\
    \sinh\rho_0 \sin {j\pi \over 2} \end{pmatrix} \,.
 }
Let the initial velocities be
 \eqn{vSquare}{
  v_j = \begin{pmatrix} 0 \\ 1 \\ -(-1)^j \sin {j\pi \over 2} \\
    (-1)^j \cos {j\pi \over 2} \end{pmatrix} \,.
 }
The first collision after time $\tau=0$ occurs at a time $\tau = \Delta\tau/2$ where
 \eqn{DeltaTauDef}{
  \Delta\tau = 2 \arctan\tanh\rho_0 \,.
 }
At this time, one can easily see that
 \eqn{hCollide}{
  h_1 = h_4 = \begin{pmatrix} \cosh\rho_0 \\ \sinh\rho_0 \\
   \sinh\rho_0 \\ \sinh\rho_0 \end{pmatrix} \qquad\qquad
  h_2 = h_3 = \begin{pmatrix} \cosh\rho_0 \\ \sinh\rho_0 \\
   -\sinh\rho_0 \\ -\sinh\rho_0 \end{pmatrix} \,.
 }
One can check that after an additional interval of time $\Delta\tau/2$, the string is again in the form of a perfect square, and after another such interval there is another double collision, with the string orthogonal to its configuration as indicated in \eno{hCollide}.

Now consider the energy of this string at $\tau=0$.  Because of symmetry, we can look at only one side of the square and then multiply the result by $4$.  To proceed with the segment between kink $4$ and kink $1$, we first need to observe that the vector $k$ that defines the $AdS_2$ face along which this segment runs is
 \eqn{kSquare}{
  k = \begin{pmatrix} \sinh\rho_0 \\ \cosh\rho_0 \\ \cosh\rho_0 \\ \cosh\rho_0
       \end{pmatrix} \,,
 }
from which we immediately extract
 \eqn{tkmValues}{
  \theta = {\pi \over 4} \qquad
  \kappa = \arcsinh(\sqrt{2} \cosh\rho_0) \qquad
  \mu = \arctan\coth\rho_0 \,.
 }
Starting from \eno{PttComplicated}-\eno{EComplicated}, one can show that the total energy is
 \eqn{EtotSquare}{
  E_{\rm tot} = 4 E_{41} = {4 \sinh 2\rho_0 \over 2\pi\alpha'} \,.
 }
As a spot-check of our calculations in the previous section, one can re-evaluate the energy at time $\tau = -\Delta\tau/2$.  This is a convenient time to choose because the string spans its widest extent on the $AdS_2$ face described by \eno{kSquare}.  After rotating coordinates as described above \eno{SpecialCoords}, the extent of the string is from $Y^1 = -\sqrt{2} \sinh\rho_0$ to $Y^1 = \sqrt{2} \sinh\rho_0$.  Applying \eno{PttSpecial}-\eno{ESpecial} with $\tau = -\Delta\tau/2$, one swiftly arrives at
 \eqn{EtSagain}{
  E_{\rm tot} = 2 E_{41} = {4 \sinh 2\rho_0 \over 2\pi\alpha'} \,,
 }
where the first equality follows from noting that at time $\tau = -\Delta\tau/2$, the string runs from $h_0$ to $h_1$ and then doubles back on itself to run back from $h_2=h_1$ to $h_3=h_4$.  The agreement between the final expressions in \eno{EtotSquare} and \eno{EtSagain} is a consequence of energy conservation and serves as the desired spot-check.

\subsection{Semi-classical analysis of yo-yo strings in $AdS_5$}
\label{WKB}

In \cite{Gubser:2002tv}, long folded strings were considered which spin rigidly in global $AdS_5$, and which are argued to be dual to operators in ${\cal N}=4$ super-Yang-Mills theory involving many gauge-covariant derivatives, such as $\tr X^I \nabla_{(\mu_1} \cdots \nabla_{\mu_S)} X^I$.  Here $(\cdots)$ indicates traceless symmetrization so as to obtain a spin-$S$ representation.  The main calculation focuses on a string state whose dual operator is more precisely described as $\tr X^I \nabla_z^S X^I$ where $\nabla_z = \nabla_2 - i \nabla_3$, and $\nabla_2$ and $\nabla_3$ are understood as covariant derivatives in two chosen spatial directions.  The energy $E$ of the string state in global $AdS_5$ (rendered dimensionless by a factor of the $AdS_5$ radius $L$) is interpreted as the dimension of the dual operator, while the angular momentum of the string state is just $S$.  An interesting expression for the twist,
 \eqn{LargeTwist}{
  \Delta - S = {\sqrt\lambda \over \pi} \log {S \over \sqrt\lambda} + {\cal O}(S^0) \,,
 }
was recognized as relating to the cusp anomalous dimension of Wilson loops.  A significant fraction of the integrability literature has been devoted to expanded understanding of this type of string / operator mapping, together with the field theory analysis of the field theory operators.  In this section, we would like to pursue the semi-classical quantization of the yo-yo string in order to probe its possible relation to operators similar to the ones that describe the folded spinning string.  For a related calculation based on a different classical string motion, see \cite{Minahan:2002rc}.

So far we have considered the yo-yo string in $AdS_3$, but the generalization to any dimension of anti-de Sitter space is obvious; in AdS of any dimension, the string worldsheet still stays within an $AdS_2$ subspace.  We will focus on $AdS_5$.  Following the presentation of \cite{Ficnar:2013wba}, we write an action
 \eqn{ActionWithEndpoint}{
  S = -{1 \over 4\pi\alpha'} \int_M d^2 \sigma \, \sqrt{-h} h^{ab}
    \partial_a X^\mu \partial_b X^\nu G_{\mu\nu} + 
   \int_{\partial M} d\xi \, {1 \over 2\eta} \dot{X}^\mu \dot{X}^\nu G_{\mu\nu} \,, 
 }
where the second term is included to describe momentum at the endpoints, since in this section we are considering a single $AdS_5$ yo-yo.  Here $h_{ab}$ is the worldsheet metric, determined up to a conformal factor by its equation of motion, and $\eta = \eta(\xi)$ is the einbein on the boundary, whose choice is equivalent to choosing a particular coordinate $\xi$ to parametrize the boundary.  We will consider motions in global $AdS_5$, described as
 \eqn{GlobalAdS}{
  ds^2 = L^2 (-\cosh^2 \rho \, dt^2 + d\rho^2 + \sinh^2 \rho \, d\Omega_3^2) \,,
 }
where $d\Omega_3^2$ is the metric on a unit $S^3$.  For simplicity, let us set the $AdS_5$ radius $L=1$.

To treat the folded string efficiently, we use $t$ and $\rho$ to parametrize the worldsheet; only $\rho$ runs from $0$ to $\infty$, such that this parametrizes only half of the worldsheet of the yo-yo centered around the origin. We thus integrate over only half the worldsheet, but will then double the resulting action.  Also, we parametrize the boundary using $\xi=t$; its location will be denoted $\rho_*(t)$.  The action can be rewritten as
 \eqn{ActionStatic}{
  S = \int dt \, {\cal L}
 }
where
 \eqn{LagrangianStatic}{
  {\cal L} = -{2 \over \pi\alpha'} \int_0^{\rho_*} d\rho \, \cosh\rho + 
    {1 \over \eta} (-\cosh^2 \rho_* + \dot\rho_*^2)
   = -{2 \over \pi\alpha'} \sinh\rho_* + 
       {1 \over \eta} (-\cosh^2 \rho_* + \dot\rho_*^2) \,.
 }
The coefficient $-2/\pi\alpha'$ on the first term comes from the coefficient $-1/4\pi\alpha'$ on the first term of \eno{ActionWithEndpoint}.  One factor of $2$ comes from plugging in the worldsheet metric for $h_{ab}$; another comes from the fact that the string is doubled over; and a third comes from the fact that we only parametrized the $\rho>0$ half of the string.  Likewise the $1/\eta$ coefficient comes from $1/2\eta$ in \eno{ActionWithEndpoint}, doubled because our parametrization only tracks one of the two endpoints.

We will now simplify notation by replacing $\rho_*$ by $\rho$.  We observe that the equation of motion for $\eta$ simply enforces that the endpoint should move along a null trajectory.  We now form the Hamiltonian
 \eqn{HamiltonianStatic}{
  H &= p\dot\rho - {\cal L}  \cr
    &= {1 \over \eta} (\dot\rho \mp \cosh\rho)^2 \pm {2 \over \eta} \dot\rho \cosh\rho + 
      {2 \over \pi\alpha'} \sinh\rho  \cr
    &= |p| \cosh\rho + {2 \over \pi\alpha'} \sinh\rho \,.
 }
In the last line we have used the null trajectory condition on the endpoint.  Alternatively, we could form $H$ as the integral of $P_t^t$ plus endpoint contributions.

The WKB condition, used to describe a quarter of a full cycle of the motion in which the endpoint of the string starts at $\rho=0$ and proceeds to its maximum value $\rho_0$, reads
 \eqn{WKBcondition}{
  {\pi \over 2} N = \int_0^{\rho_0} d\rho \, p(\rho)
 }
where $N$ is the excitation level and $p(\rho)$ is obtained by solving the equation $H = E$.  Thus
 \eqn{WKBagain}{
  {\pi \over 2} N = 
    \int_0^{\rho_0} d\rho \, \sech \rho \left( E - {2 \over \pi\alpha'} \sinh\rho \right) \,. 
 }
We determine $\rho_0$ by setting $p(\rho_0) = 0$.  When $\rho_0$ is large, we readily find
 \eqn{EminusN}{
  E - N = {4 \over \pi^2 \alpha'} \log N + {\cal O}(N^0) \,.
 }
The simplest explanation of the logarithmic term in \eno{EminusN} is that the yo-yo strings are mapped to operators which are similar to the $\tr X^I \nabla_z^S X^I$ operators of \cite{Gubser:2002tv}, where $\nabla_z = {1 \over 2} (\nabla_2 - i \nabla_3)$ and $\nabla_i$ for $i=1,2,3$ are the gauge-covariant derivatives in the three spatial directions of the boundary theory.  Specifically, consider the operator $\tr X^I \nabla_1^N X^I$.  This operator does not transform in a definite representation of the rotation group, but it overlaps with representations with spin up to $N$ and is annihilated by rotations which preserve the direction of the $1$ axis.  Heuristically, then, it is a good candidate to be mapped to the yo-yo string, which also is invariant under an abelian subgroup of rotations but can be understood to have a large total angular momentum.  A striking aspect of \eno{EminusN} is that the coefficient of $\log N$ is $4/\pi$ times the result of \eno{LargeTwist}, and we do not have a clear account for why this factor should be present.

\section{Further examples}
\label{EXAMPLES}

We now discuss four example evolutions of segmented strings. The first and easiest example is the square; then we deal
 with the regular hexagon and the regular octagon, followed 
by an example of irregular shape, namely an irregular hexagon.

We have written an algorithm in Mathematica that iterates the collisions. In particular we focus our attention on the analysis of the periodicity of the motion and on the conservation of the energy of the aforementioned systems of strings of various shapes. Anticipating our results, we find that the motion is periodic only in the case of the square. For other regular and irregular shapes taken into consideration the motion does not repeat itself periodically, at least not on the timescales we analyzed. Nevertheless, we verified that the energy is conserved in all cases, and this constitutes a nontrivial check of the correctness of the algorithm we used in the evolution of the motion. 

As discussed in section \ref{icSubsection}, for each considered shape of an even number ($N$) of kinks we will need to specify $\vec{h}_1,\vec{h}_2,\ldots,\vec{h}_{N-1}$ and $\vec{v_1}$, in order to be able to fully determine the initial conditions. For the regular $N$-gon, this comes down to specifying the $N$-gon radius $r$, offset angle $\phi$ (defined as the angle the first vertex makes with the $Y^1$-axis), sign $\sigma$ and magnitude of velocity $v$ in the general formulas    
\begin{equation}\label{NGONh}
h_N = \left( \sqrt{1+r^2}, \, 0,\, r \cos \left( (i-1)\frac{2\pi}{N} + \phi \right),\, r \sin \left( (i-1)\frac{2\pi}{N} + \phi \right) \right), 
\end{equation}
\begin{equation}\label{NGONv}
v_N = \left( 0,\, v,\,  -\sigma (-1)^i v \sin \left( (i-1)\frac{2\pi}{N} + \phi \right),\,  \sigma (-1)^i v  \cos \left( (i-1) \frac{2\pi}{N} + \phi \right) \right)
\end{equation}
with $i=1, \ldots, N$.

In addition to snapshots of the string configurations shown below, we
provide videos of the evolution of the string for all our examples
online, see \cite{website}.

\clearpage
\subsection{Square}
\label{sec:square}

As discussed earlier in section \ref{exampleSquare}, the motion of the square is periodic, with a time interval $\Delta \tau$ between collisions and a conserved energy $E$ that can both be calculated analytically. We performed a check on this by running the program for a square ($N=4$) with initial conditions specified by choosing $r=\sqrt 2$, $\phi = \pi/4$, $\sigma=1$ and $v = r = \sqrt 2$ in formulas \eqref{NGONh}-\eqref{NGONv}. See \cite{vRegSquare} for the video of the evolution of the configuration. 
Due to the symmetry of the initial string configuration, two collisions of neighboring vertices always happen simultaneously. 
We have tracked the motion for 60 collision events, \textit{i.\,e.}, for the first $120$ collisions of vertex pairs. We observe that the events are equally spaced in AdS$_3$ time, with
\begin{equation}
\Delta \tau = 1.36944 \,.
\end{equation}
This means that each vertex undergoes collisions with a period of $\Delta\tau$.  
The energy is conserved at a value of $E =  \frac{1}{2 \pi \alpha'} \times 19.5959$.  It is easy to see that these values agree with \eno{DeltaTauDef} and \eno{EtotSquare}.

We will see in the next examples that the string trajectories are no longer periodic once $N>4$, even when we deal with regular shapes.

\clearpage
\subsection{Regular hexagon}
\label{sec:regular-hexagon}

The initial $h$ and $v$ for the regular hexagon ($N=6$) are shown in Table \ref{myTablehv}. 
\begin{table}[h!]
    \caption{Initial values of vectors $h_i$ and $v_i$ for the regular hexagon.\label{myTablehv}}
    \begin{minipage}{.5\linewidth}
    \centering
              \begin{tabular}{rcccccc}
\toprule
 & $h_1$ & $h_2$ & $h_3$ & $h_4$ & $h_5$ & $h_6$ \\
\cmidrule{2-7}
$-1$\phantom{x;} & $ \frac{\sqrt7}{2} $ & $ \frac{\sqrt7}{2} $ & $ \frac{\sqrt7}{2} $ & $ \frac{\sqrt7}{2} $ & $ \frac{\sqrt7}{2} $ & $\frac{\sqrt7}{2} $ \\\addlinespace[.6ex]
   0\phantom{x;} & 0 & 0 & 0& 0 &0 &0 \\\addlinespace[.6ex]
   1\phantom{x;} & $\frac34 $ & 0 & $ -\frac34 $ & $ -\frac34 $ & 0 & $ \frac34 $ \\\addlinespace[.6ex]
   2\phantom{x;} & $ \frac{\sqrt3}{4} $  &$ \frac{\sqrt3}{2} $ & $ \frac{\sqrt3}{4} $ & $ \frac{\sqrt3}{4} $ & $ \frac{\sqrt3}{2} $ &  $- \frac{\sqrt3}{4} $ \\
\bottomrule
\end{tabular}
    \end{minipage}%
    \begin{minipage}{.5\linewidth}
    \centering
 \begin{tabular}{rcccccc}
\toprule
 & $v_1$ & $v_2$ & $v_3$ & $v_4$ & $v_5$ & $v_6$ \\
\cmidrule{2-7}
$-1$\phantom{x;} & 0 & 0 & 0 & 0& 0 &0  \\\addlinespace[.6ex]
   0\phantom{x;} & 1 & 1 & 1 & 1  & 1 & 1 \\\addlinespace[.6ex]
   1\phantom{x;} & $-\frac12 $ & 1 & $ -\frac12 $ & $ -\frac12 $ & 1 & $ -\frac12 $ \\\addlinespace[.6ex]
   2\phantom{x;} & $ \frac{\sqrt3}{2} $  &$ 0 $ & $ -\frac{\sqrt3}{2} $ & $ \frac{\sqrt3}{2} $ & 0 &  $- \frac{\sqrt3}{2} $ \\
\bottomrule
\end{tabular}
    \end{minipage} 
\end{table}
\par\noindent
They follow from setting $r=\sqrt 3/2$, $\phi = \pi/6$, $\sigma=-1$ and $v = 1$ in formulas \eqref{NGONh}--\eqref{NGONv}. 

\begin{figure}[p]
  \centering
  {\includegraphics[height=49mm]{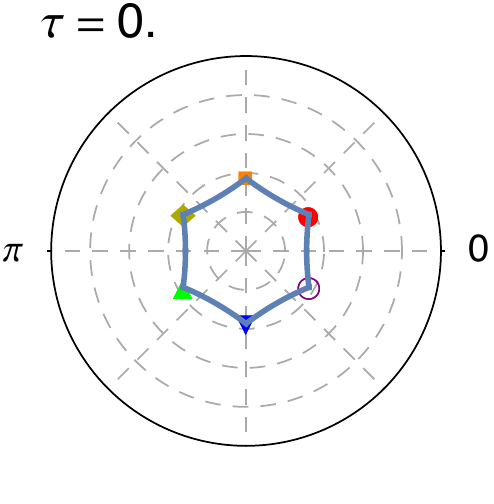}}
  {\includegraphics[height=49mm]{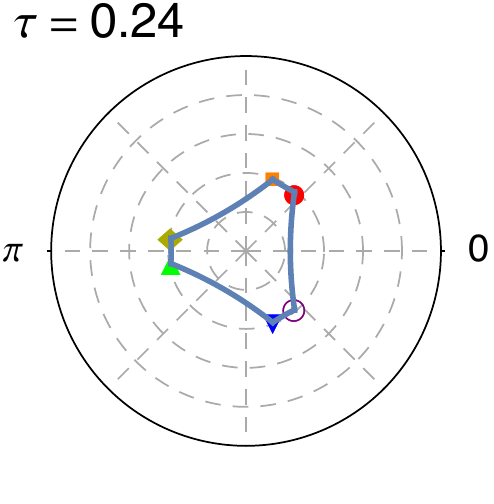}}
  {\includegraphics[height=49mm]{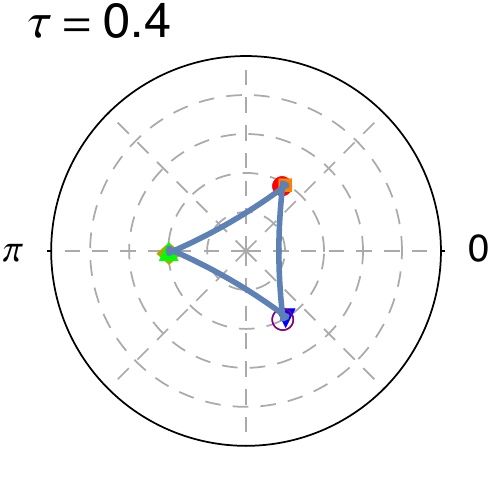}}
  {\includegraphics[height=49mm]{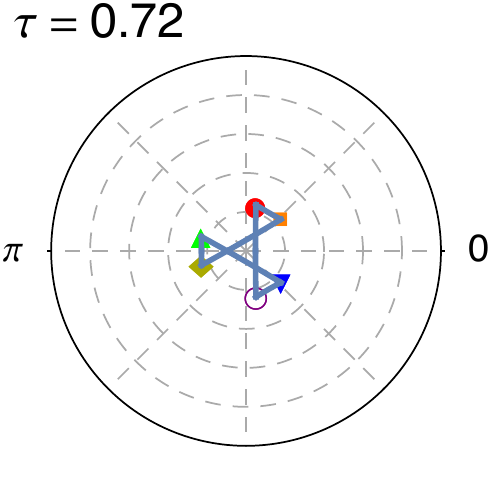}}
  {\includegraphics[height=49mm]{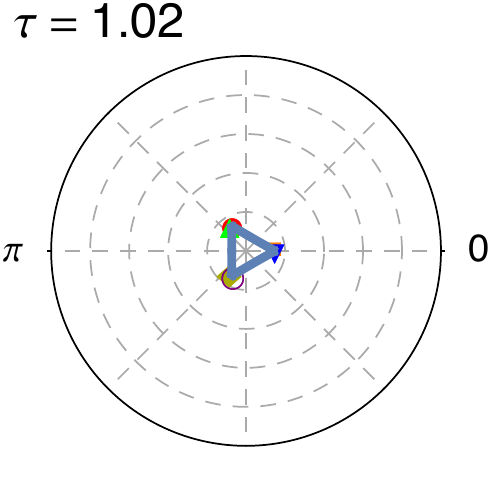}}
  {\includegraphics[height=49mm]{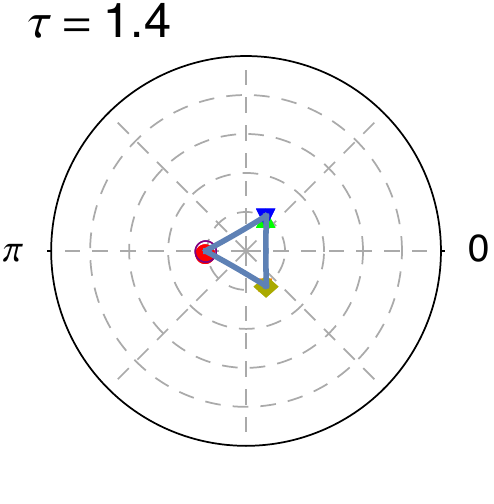}}
  {\includegraphics[height=49mm]{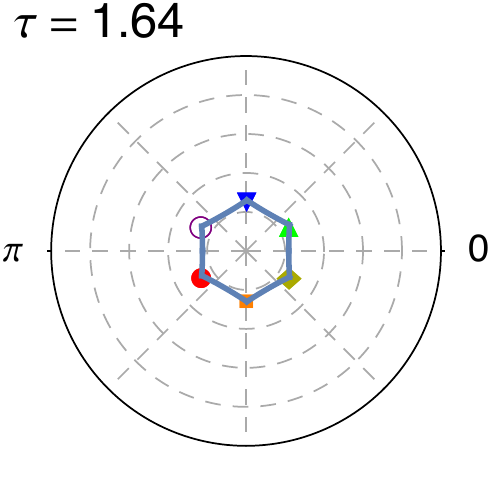}}
  {\includegraphics[height=49mm]{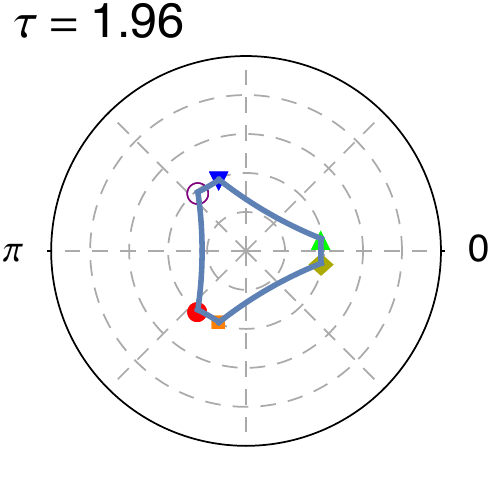}}
  {\includegraphics[height=49mm]{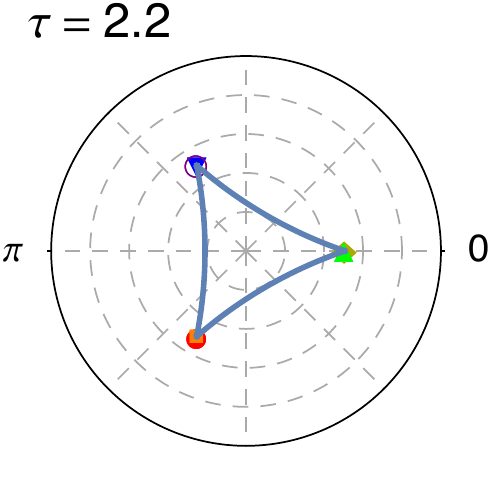}}
  \caption{Pattern of motion of the regular hexagon for
    $0\le\tau\le 2.2$. The string configuration is projected onto the
    Poincar\'e disk. Due to the regularity of the initial
      conditions, all vertex-pair collisions always happen
      simultaneously. The plots in the right column show the first
    nine ($=3\times3$) collisions.}
  \label{reg_hex_figure}
\end{figure}

\begin{figure}[p]
  \begin{subfigure}[b]{\textwidth}
    \centering
    \includegraphics{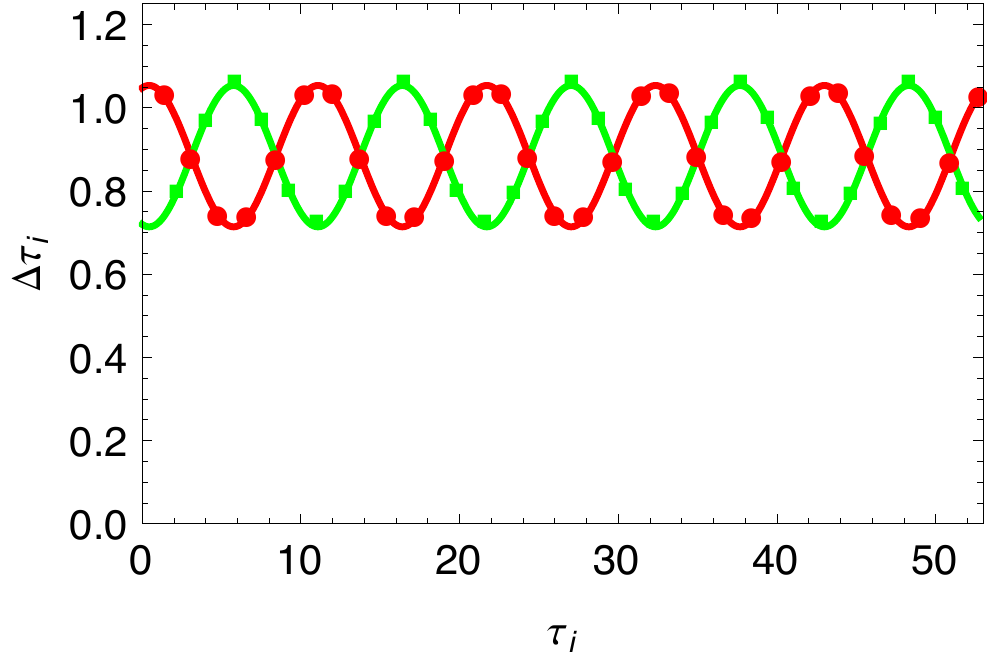}
    \caption{Collision time intervals $\Delta\tau_i$ in AdS units,
      plotted versus the time $\tau_i$
        at the upper end of the respective interval. At every data point three collisions
      happen simultaneously. The solid lines are fits of
      the form $a+b\sin(\omega\tau_i+\phi)$.}
    \label{fig:dtau_fit_reg_hex}
  \end{subfigure}\\[1em]
  \begin{subfigure}[b]{\textwidth}
    \centering
    \includegraphics{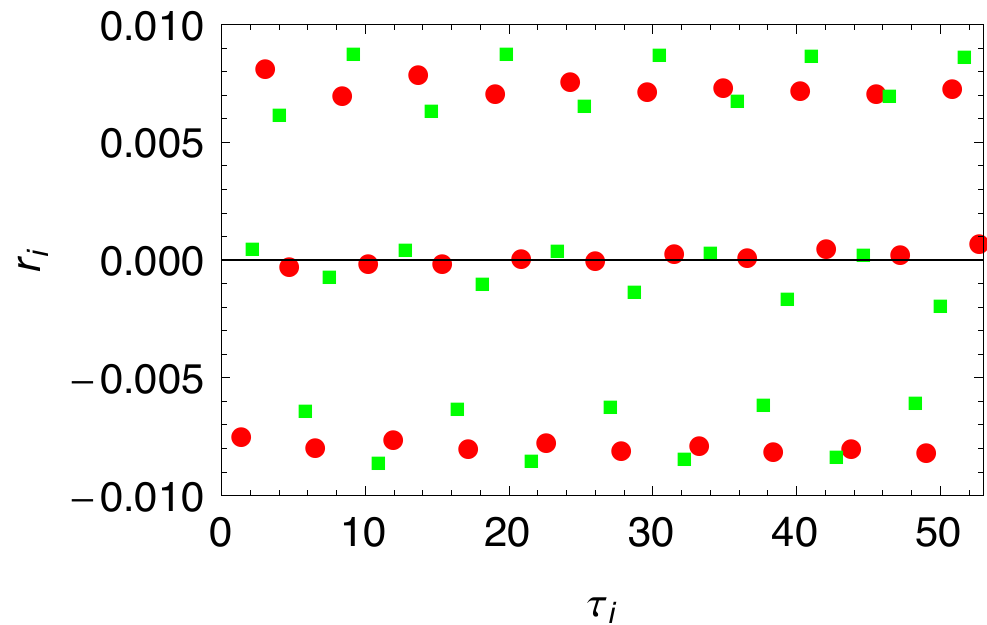}\hspace*{1.4em}
    \caption{Residues of the fit shown in the upper panel, plotted
      versus the time $\tau_i$ of collision.}
    \label{fig:res_reg_hex}
  \end{subfigure}
  \caption{Collision time intervals for the regular
    hexagon. Apparently, the data points can, alternatingly, be
    described by shifted sines (upper panel), with residues of
    order $10^{-2}$ (lower panel).}
  \label{fig:reg_hex_fit_and_res}
\end{figure}

We let the system evolve and find a more elaborate pattern for the motion of the strings. The sequence in Fig.~\ref{reg_hex_figure} represents 
the motion from the beginning until the third set of collisions. Due to the symmetry of the initial configuration, all pairs of neighboring vertices always collide simultaneously. In other words, every set of collisions comprises three collisions of vertex pairs. A video of the evolution of the motion for the hexagon can be found in \cite{vRegHex}.  Conservation of energy is a useful check on the numerics.  We find that
\begin{equation}
E = \frac{1}{2 \pi \alpha'} \times 7.93725
\end{equation}
is indeed constant throughout the motion.  The motion does not appear to be periodic, but a sort of quasi-periodicity is evident from the plot in Fig.~\ref{fig:reg_hex_fit_and_res} of the first $59$ intervals $\Delta\tau_i$ between collisions against the times $\tau_i$ at which they occur.  We will characterize this quasi-periodicity more precisely in section~\ref{sec:patt-coll-time}.
\clearpage

\clearpage
\subsection{Regular octagon}
\label{sec:regular-octagon}

\begin{figure}[p]
  \begin{subfigure}[b]{\textwidth}
    \centering
    \includegraphics{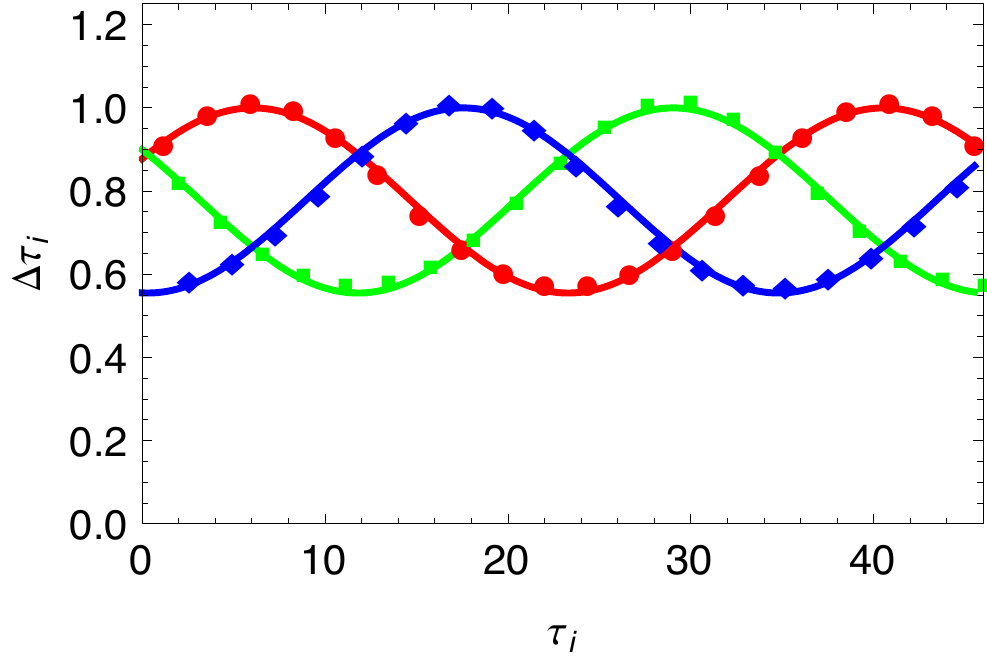}
    \caption{Collision time intervals $\Delta\tau_i$ in AdS units,
      plotted versus the time $\tau_i$
        at the upper end of the respective interval.
        At  every data point four collisions
       happen simultaneously. The solid lines are fits of
      the form $a+b\sin(\omega \tau_i+\phi)$.}
    \label{fig:dtau_fit_reg_oct}
  \end{subfigure}\\[1em]
  \begin{subfigure}[b]{\textwidth}
    \centering
    \includegraphics{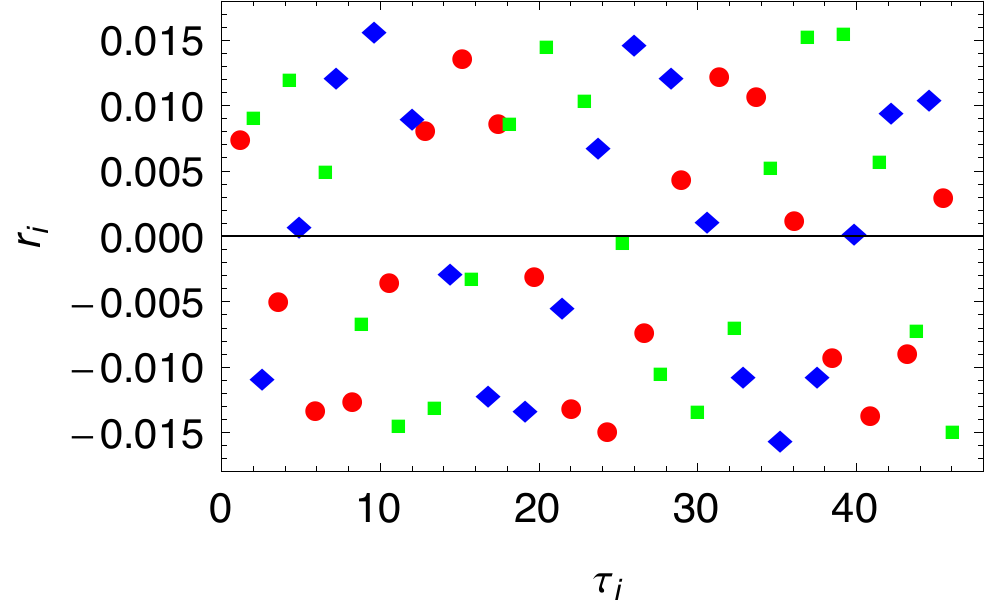}\hspace*{1.4em}
    \caption{Residues of the fit shown in the upper panel, plotted
      versus the times $\tau_i$ of collision.}
    \label{fig:res_reg_oct}
  \end{subfigure}
  \caption{Collision time intervals for the regular
    octagon. Apparently, the data points can, alternatingly, be
    described by shifted sines (upper panel), with residues of
    order  $10^{-2}$ (lower panel).}
  \label{fig:reg_oct_fit_and_res}
\end{figure}

We constructed the initial conditions for the regular octagon starting from formulas \eqref{NGONh} and \eqref{NGONv} with $N=8$, $r=1$, $\phi = \pi/4$, $\sigma=-1$ and $v = r = 1$ (with the resulting initial $h$ and $v$ shown in Table \ref{myTablehvOct} for completeness). The video of the motion of the regular octagon configuration can be found in \cite{vRegOct}.  Similar to the case of the regular hexagon, four vertex-pair collisions always happen simultaneously.  The energy is conserved, as it must be, assuming the constant value
\begin{equation}
E= \frac{1}{2 \pi \alpha'} \times 9.37258 
\end{equation}
throughout the motion.  As before, the motion is not periodic, but quasi-periodicity can be observed in the plot in Fig.~\ref{fig:reg_oct_fit_and_res} of the first $59$ intervals $\Delta\tau_i$ between collisions.  

\begin{table}[h!]
    \caption{Initial values of vectors $h_i$ and $v_i$ for the regular octagon.\label{myTablehvOct}}
    \centering
              \begin{tabular}{rcccccccc}
\toprule
 & $h_1$ & $h_2$ & $h_3$ & $h_4$ & $h_5$ & $h_6$ & $h_7 $ &$  h_8$\\
\cmidrule{2-9}
$-1$\phantom{x;} & $ \sqrt2 $ & $ \sqrt2 $ & $ \sqrt2 $ & $ \sqrt2 $ & $ \sqrt2 $ & $ \sqrt2 $ & $ \sqrt2 $ & $\sqrt2$ \\\addlinespace[.6ex]
   0\phantom{x;} & 0 & 0 & 0& 0 &0 &0 &0&0\\\addlinespace[.6ex]
   1\phantom{x;} & $\frac{1}{\sqrt2} $ & 0 & $ -\frac{1}{\sqrt2} $ & $ -1 $ & $ -\frac{1}{\sqrt2} $ & $ 0 $ & $ \frac{1}{\sqrt2} $ & $1$ \\\addlinespace[.6ex]
   2\phantom{x;} & $ \frac{1}{\sqrt2} $  &$ 1 $ & $ \frac{1}{\sqrt2} $ & $ 0 $ & $ -\frac{1}{\sqrt2} $ &  $- 1 $ & $ -\frac{1}{\sqrt2} $ & $ 0$  \\
\midrule
 & $v_1$ & $v_2$ & $v_3$ & $v_4$ & $v_5$ & $v_6$ & $v_7$ & $v_8$ \\
\cmidrule{2-9}
$-1$\phantom{x;} & 0 & 0 & 0 & 0& 0 &0 & 0&0 \\\addlinespace[.6ex]
   0\phantom{x;} & 1 & 1 & 1 & 1  & 1 & 1 &1 &1 \\\addlinespace[.6ex]
   1\phantom{x;} & $-\frac{1}{\sqrt2} $ & 1 & $ -\frac{1}{\sqrt2} $ & $ 0 $ & $\frac{1}{\sqrt2} $ & $ -1 $ &  $ \frac{1}{\sqrt2} $ &  $ 0$  \\\addlinespace[.6ex]
   2\phantom{x;} & $ \frac{1}{\sqrt2} $  &$ 0 $ & $ -\frac{1}{\sqrt2} $ & $ 1 $ & $-\frac{1}{\sqrt2}$ &  $ 0 $ & $ \frac{1}{\sqrt2}$ & $ -1 $ \\
\bottomrule
\end{tabular}
\end{table}

\clearpage
\subsection{Irregular hexagon}
\label{sec:irregular-hexagon}

For the irregular hexagon, the spatial values of the positions $h_N$ and velocity $v_1$ can be specified more arbitrarily. Table \ref{myTablehvirreg} shows the numbers we used as input to run the program (with finite precision in this case).
\begin{table}[h!]
    \caption{Initial values of vectors $h_i$ and $v_i$ for the irregular hexagon.\label{myTablehvirreg}}
    \centering
              \begin{tabular}{rcccccc}
\toprule
 & $h_1$ & $h_2$ & $h_3$ & $h_4$ & $h_5$ & $h_6$ \\
\cmidrule{2-7}
$-1$\phantom{x;} &  1.40046  &   1.40046  &  1.2929   &   1.2929  & 1.35095   &  1.35095  \\\addlinespace[.3ex]
   0\phantom{x;} & 0 & 0 & 0& 0 &0 &0 \\\addlinespace[.3ex]
   1\phantom{x;} & 0.845877 & 0.00640535 & $-0.715193$ & $-0.715193$ & 0.0119017 & 0.780614 \\\addlinespace[.3ex]
   2\phantom{x;} & 0.495763 & 0.980432 & 0.400119 & $-0.400119$ & $-0.908246$ & $-0.46443$  \\
\midrule
 & $v_1$ & $v_2$ & $v_3$ & $v_4$ & $v_5$ & $v_6$ \\
\cmidrule{2-7}
$-1$\phantom{x;} & 0.187146 &  0.187146 &  $-0.0467448$ & $-0.0467448$ & 0.0674737 & 0.0674737 \\\addlinespace[.3ex]
   0\phantom{x;} & 1.16403 & 1.16403 & 0.952793 & 0.952793 & 1.0514 & 1.0514 \\\addlinespace[.3ex]
   1\phantom{x;} & $-0.35$ &  1.15 & $-0.4$ & $-0.4$ & 1.05 & $-0.45$ \\\addlinespace[.3ex]
   2\phantom{x;} & 1.12583 & 0.259808 & $-0.866025$ & 0.866025 & $-0.0866025$ &  $-0.952628$  \\
\bottomrule
\end{tabular}
\end{table}
\par
\begin{figure}[p]
  \centering
  {\includegraphics[height=49mm]{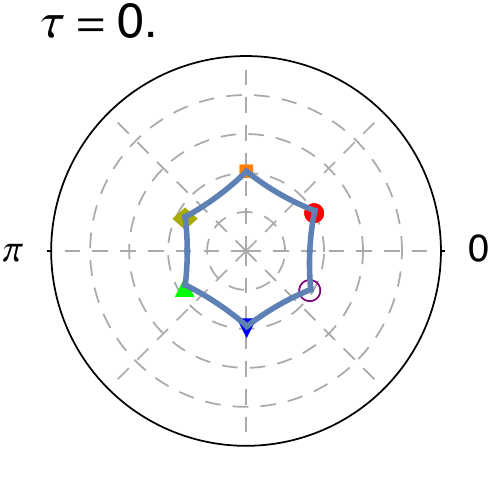}}
  {\includegraphics[height=49mm]{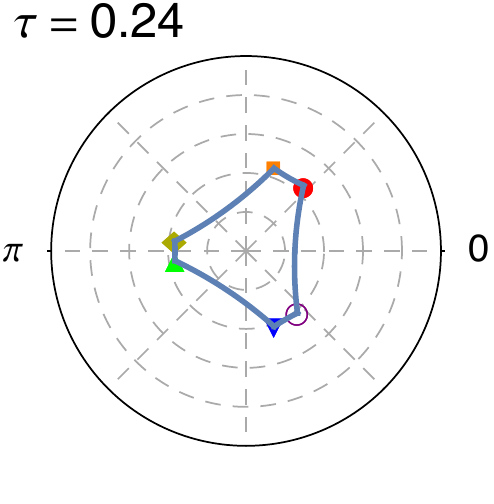}}
  {\includegraphics[height=49mm]{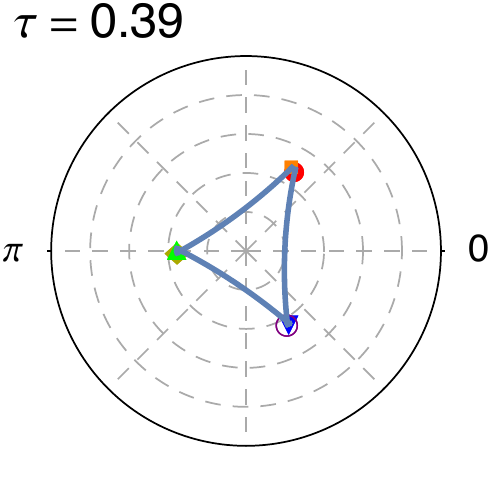}}
  {\includegraphics[height=49mm]{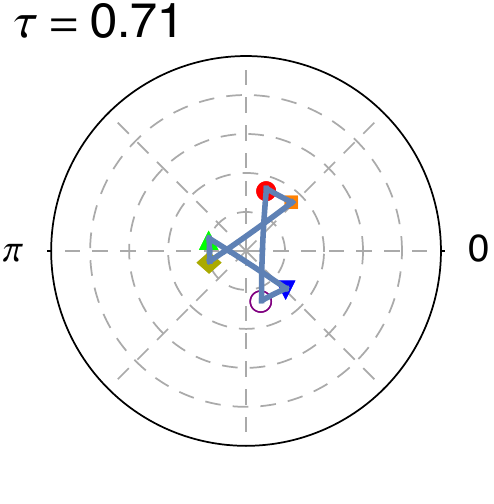}}
  {\includegraphics[height=49mm]{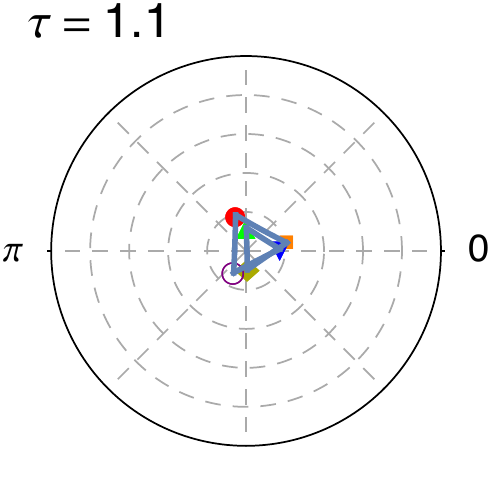}}
  {\includegraphics[height=49mm]{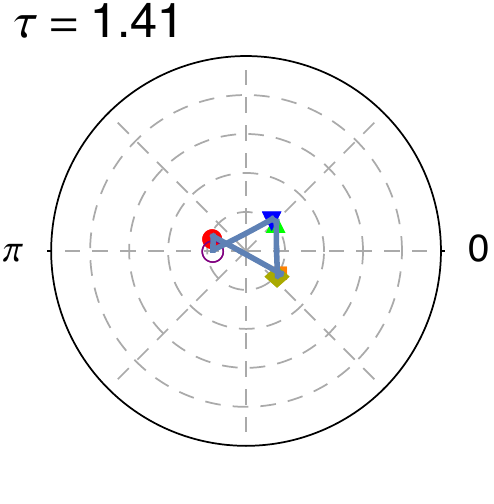}}
  {\includegraphics[height=49mm]{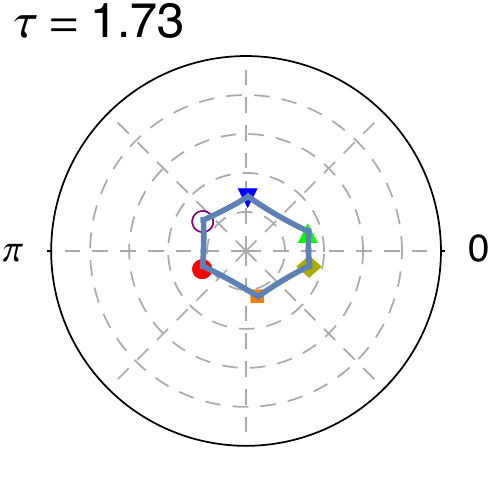}}
  {\includegraphics[height=49mm]{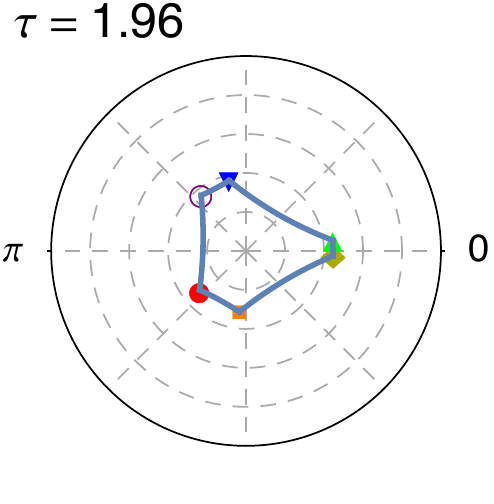}}
  {\includegraphics[height=49mm]{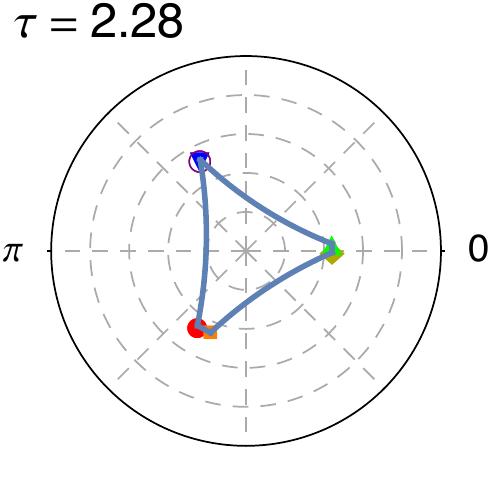}}
  \caption{Pattern of motion for the irregular hexagon for
    $0\le\tau\le 2.28$. The string configuration is projected onto the
    Poincar\'e disk. As opposed to the case of the regular hexagon
    and octagon, collisions of vertex pairs do not happen at the same time. The plots show the string motion up to the ninth collision.}
  \label{irreg_hex_figure}
\end{figure}

\begin{figure}[t]
    \centering
    \includegraphics[width=0.5\textwidth]{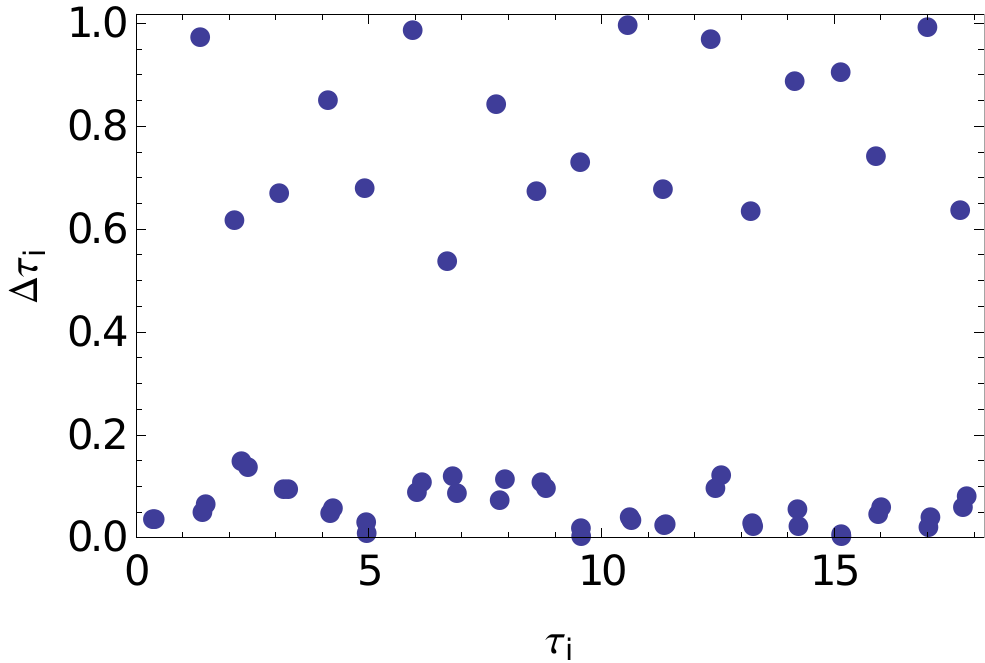}\caption{Collision time intervals $\Delta\tau_i$ in AdS units for the irregular hexagon, plotted versus the time $\tau_i$
        at the upper end of the respective interval.}
\label{irregdeltatau}
\end{figure}

\noindent
The resulting irregular motion is depicted in Fig \ref{irreg_hex_figure} and shown in the video \cite{vIrregHex}.  As in the other examples considered, the irregular hexagon passes the non-trivial check that energy is conserved during the evolution of the vertices, at a value of 
\begin{equation}
E= \frac{1}{2 \pi \alpha'} \times 8.56488.  
\end{equation}
The pattern of collision time intervals, shown in Fig \ref{irregdeltatau}, is irregular, as expected. No periodic structure can be distinguished.

\clearpage
\subsection{Patterns in the collision time intervals}
\label{sec:patt-coll-time}

For the regular $N$-gons with $N>4$ that we consider, \textit{viz.}
the regular hexagon and octagon in sections \ref{sec:regular-hexagon}
and \ref{sec:regular-octagon}, respectively, we found that the motion
is not periodic in time. Quite intriguingly, however, the time
intervals between subsequent collisions can be fitted with shifted
sine functions.
In the case of the hexagon, the collision time intervals $\Delta\tau_i$
appear to alternatingly follow one of two sines, \textit{cf.}
Fig.~\ref{fig:reg_hex_fit_and_res}, whereas for the octagon there are
three sines, \textit{cf.}  Fig.~\ref{fig:reg_oct_fit_and_res}.
In each of the cases, the different oscillations that we discern share
the same frequency to high accuracy.

More precisely, we were able to fit the numerical data for the hexagon to the form
 \eqn{HexagonForm}{
  \Delta\tau_i = \left\{ \seqalign{\span\TR & \qquad\span\TT}{
   a + b \sin( \omega \tau_i + \phi) & for odd $i$  \cr
   a + b \sin( \omega \tau_i + \phi + \pi ) & for even $i$} \right.
 }
and the best fit parameters are
 \eqn{HexagonValues}{
  a = 0.883603 \,, \qquad b = 0.169865 \,, \qquad \omega = 0.590824 \,, \qquad
    \phi = 1.30763 \,.
 }
The residuals of the fit are on order $10^{-2}$, which is small but numerically significant: see the lower panel of Fig.~\ref{fig:reg_hex_fit_and_res}.  Sinusoidal structure can be detected in the residuals.

The numerical data for the octagon can similarly be fit to the form
 \eqn{OctagonForm}{
  \Delta\tau_i = \left\{ \seqalign{\span\TR & \qquad\span\TT}{
   a + b \sin( \omega \tau_i + \phi) & for $i \equiv 1 \ ({\rm mod}\ 3)$  \cr
   a + b \sin( \omega \tau_i + \phi + 2\pi/3 ) & for $i \equiv 2 \ ({\rm mod}\ 3)$  \cr
   a + b \sin( \omega \tau_i + \phi + 4\pi/3 ) & for $i \equiv 0 \ ({\rm mod}\ 3)$
   } \right.
 }
and the best fit parameters are
 \eqn{OctagonValues}{
  a = 0.777052 \,, \qquad b = 0.222491 \,, \qquad \omega = 0.182199 \,, \qquad
    \phi = 0.466574
 }
with slightly larger residuals than in the case of the hexagon.  As before, sinusoidal behavior can be detected in the residuals: see the lower panel of Fig.~\ref{fig:reg_oct_fit_and_res}.

\clearpage

\section{Conclusions}

We have explained segmented string solutions in flat space and in $AdS_3$, where in a given time snapshot each segment is in the flat space case a straight line, or in the $AdS_3$ case the intersection of an $AdS_2$ subspace and a surface of fixed global time.  Tracking the motion of these segments is easy until their endpoints collide.  As we have argued, the outcome of such collisions is in fact straightforward to predict based on considerations that can be phrased entirely locally, in terms of information arbitrarily close to the collision.  The result is a pleasingly sparse specification of classical string motions, which however are exact solutions of the string equations of motion.

In flat space, all classical string motions with finite energy are periodic up to an overall motion of the center of mass, and of course segmented strings inherit this property.  In $AdS_3$, the simplest segmented motions are periodic, but less simple ones are not, at least as far as we can tell.  Instead we have numerical hints of a notion of quasi-periodicity, in which the time between vertex collisions cycles among several periodic functions, with small residuals which may themselves have similar representations.  It seems likely that some aspect of integrability is at work, and it would clearly be appealing to find a representation of these segmented motions that makes their almost-quasi-periodic behavior manifest.

It is obvious in flat space that segmented strings can be used to approximate an arbitrary string trajectory to any desired accuracy.  The simplest argument to this effect is the one following \eno{RightLeftExpand}, namely that arbitrary $Y_R$ and $Y_L$ with null tangents can be approximated by piecewise linear $Y_R$ and $Y_L$ where each piece is null.  We have not shown that an analogous statement holds in $AdS_3$, but it seems to us likely that it does.  Is there something fundamental about approximating a classical string by a collection of yo-yo solutions bound together at their endpoints?  Is some quantum mechanical treatment available based on such a picture?  A first step toward answering the second question might be to make a more systematic study of quantum states of the yo-yo, beyond the semi-classical regime, or in some improved version of the WKB treatment that we gave.

For simplicity, we have avoided localized momentum at the kinks where segments join together.  This seems unnatural from the point of view of the previous paragraph, where we do our best to take seriously the assemblage of yo-yo solutions as a guide to the actual dynamics of strings.  We believe that localized momentum could be included in our formalism, though obviously it would complicate the treatment of kink collisions.  However, from a certain point of view it should be unnecessary.  Localized momentum at a kink can be approximated by replacing the kink by two kinks very close to one another with a string segment between them that moves nearly at the speed of light.  This claim can be demonstrated easily for specific motions. For example, a string that is doubled over on itself to form a closed string version of the yo-yo can be converted into a very thin rectangle.  A more general demonstration would be desirable.

Localized momentum presents an interesting conceptual puzzle.  What happens if we start with a very long straight string, in flat space or anti-de Sitter space, and let it collapse inward?  The localized momentum at each endpoint accumulates until it back-reacts significantly on the metric, producing some version of an Aichelburg-Sexl metric, with a string coming out one side.  When these shock waves collide, a black hole is formed.  What comes next in the evolution?  In the classical gravity picture, all that is left is a horizon, which presumably settles down to a spherical shape after some non-linear ringing.  Do the early stages of the ringing approximately follow the perturbative motion of a string re-emerging from the collision with finite momentum at its endpoints?  Or is the perturbative picture essentially lost because of the strong gravitational interactions?

\subsection*{Note added}

While this paper was in preparation, we received \cite{Vegh:2015ska}, which has some overlap with the present work.

\section*{Acknowledgments}

The work of N.C.\ was supported by a Fellowship of the Belgian American Educational Foundation.  
The work of S.S.G.\ was supported in part by the Department of Energy under Grant No.~DE-FG02-91ER40671. 
A.S.\ acknowledges support in the framework of the cooperation
contract between the GSI Helmholtzzentrum f\"ur Schwerionenforschung
and Heidelberg University.
A.S.\ thanks the Princeton Physics Department for hospitality.
The work of A.S.\ at Princeton University was supported by the
HGS-HIRe Abroad program and the ExtreMe Matter Institute EMMI.
C.T.\ acknowledges support from Columbia University and from DOE grant DE-SC0011941, and thanks the Princeton Physics Department for hospitality.

\clearpage
\bibliographystyle{ssg}
\bibliography{segmented,website}

\begingroup\raggedright\begin{thebibliography}{10}

\bibitem{Beisert:2010jr}
N.~Beisert, C.~Ahn, L.~F. Alday, Z.~Bajnok, J.~M. Drummond, {\em et.~al.},
  ``{Review of AdS/CFT Integrability: An Overview},'' {\em Lett.Math.Phys.}
  {\bf 99} (2012) 3--32, \href{http://xxx.lanl.gov/abs/1012.3982}{{\tt
  1012.3982}}.

\bibitem{Tseytlin:2010jv}
A.~A. Tseytlin, ``{Review of AdS/CFT Integrability, Chapter II.1: Classical
  AdS5xS5 string solutions},'' {\em Lett. Math. Phys.} {\bf 99} (2012)
  103--125, \href{http://xxx.lanl.gov/abs/1012.3986}{{\tt 1012.3986}}.

\bibitem{Gubser:2002tv}
S.~Gubser, I.~Klebanov, and A.~M. Polyakov, ``{A Semiclassical limit of the
  gauge / string correspondence},'' {\em Nucl.Phys.} {\bf B636} (2002) 99--114,
  \href{http://xxx.lanl.gov/abs/hep-th/0204051}{{\tt hep-th/0204051}}.

\bibitem{Arutyunov:2003rg}
G.~Arutyunov and M.~Staudacher, ``{Matching higher conserved charges for
  strings and spins},'' {\em JHEP} {\bf 03} (2004) 004,
  \href{http://xxx.lanl.gov/abs/hep-th/0310182}{{\tt hep-th/0310182}}.

\bibitem{Kruczenski:2004wg}
M.~Kruczenski, ``{Spiky strings and single trace operators in gauge
  theories},'' {\em JHEP} {\bf 08} (2005) 014,
  \href{http://xxx.lanl.gov/abs/hep-th/0410226}{{\tt hep-th/0410226}}.

\bibitem{Kazakov:2004nh}
V.~A. Kazakov and K.~Zarembo, ``{Classical / quantum integrability in
  non-compact sector of AdS/CFT},'' {\em JHEP} {\bf 10} (2004) 060,
  \href{http://xxx.lanl.gov/abs/hep-th/0410105}{{\tt hep-th/0410105}}.

\bibitem{Kalousios:2006xy}
C.~Kalousios, M.~Spradlin, and A.~Volovich, ``{Dressing the giant magnon II},''
  {\em JHEP} {\bf 03} (2007) 020,
  \href{http://xxx.lanl.gov/abs/hep-th/0611033}{{\tt hep-th/0611033}}.

\bibitem{Alday:2007hr}
L.~F. Alday and J.~M. Maldacena, ``{Gluon scattering amplitudes at strong
  coupling},'' {\em JHEP} {\bf 06} (2007) 064,
  \href{http://xxx.lanl.gov/abs/0705.0303}{{\tt 0705.0303}}.

\bibitem{Grigoriev:2007bu}
M.~Grigoriev and A.~A. Tseytlin, ``{Pohlmeyer reduction of AdS(5) x S**5
  superstring sigma model},'' {\em Nucl. Phys.} {\bf B800} (2008) 450--501,
  \href{http://xxx.lanl.gov/abs/0711.0155}{{\tt 0711.0155}}.

\bibitem{Mikhailov:2007xr}
A.~Mikhailov and S.~Schafer-Nameki, ``{Sine-Gordon-like action for the
  Superstring in AdS(5) x S**5},'' {\em JHEP} {\bf 05} (2008) 075,
  \href{http://xxx.lanl.gov/abs/0711.0195}{{\tt 0711.0195}}.

\bibitem{Jevicki:2007aa}
A.~Jevicki, K.~Jin, C.~Kalousios, and A.~Volovich, ``{Generating AdS String
  Solutions},'' {\em JHEP} {\bf 03} (2008) 032,
  \href{http://xxx.lanl.gov/abs/0712.1193}{{\tt 0712.1193}}.

\bibitem{Alday:2009yn}
L.~F. Alday and J.~Maldacena, ``{Null polygonal Wilson loops and minimal
  surfaces in Anti-de-Sitter space},'' {\em JHEP} {\bf 11} (2009) 082,
  \href{http://xxx.lanl.gov/abs/0904.0663}{{\tt 0904.0663}}.

\bibitem{Alday:2010vh}
L.~F. Alday, J.~Maldacena, A.~Sever, and P.~Vieira, ``{Y-system for Scattering
  Amplitudes},'' {\em J. Phys.} {\bf A43} (2010) 485401,
  \href{http://xxx.lanl.gov/abs/1002.2459}{{\tt 1002.2459}}.

\bibitem{Artru:1979ye}
X.~Artru, ``{Classical String Phenomenology. 1. How Strings Work},'' {\em Phys.
  Rept.} {\bf 97} (1983) 147.

\bibitem{Andersson:1983ia}
B.~Andersson, G.~Gustafson, G.~Ingelman, and T.~Sjostrand, ``{Parton
  Fragmentation and String Dynamics},'' {\em Phys.Rept.} {\bf 97} (1983)
  31--145.

\bibitem{Ficnar:2013wba}
A.~Ficnar and S.~S. Gubser, ``{Finite momentum at string endpoints},'' {\em
  Phys.Rev.} {\bf D89} (2014), no.~2 026002,
  \href{http://xxx.lanl.gov/abs/1306.6648}{{\tt 1306.6648}}.

\bibitem{Klebanov:2006jj}
I.~R. Klebanov, J.~M. Maldacena, and C.~B. Thorn, III, ``{Dynamics of flux
  tubes in large N gauge theories},'' {\em JHEP} {\bf 04} (2006) 024,
  \href{http://xxx.lanl.gov/abs/hep-th/0602255}{{\tt hep-th/0602255}}.

\bibitem{Minahan:2002rc}
J.~A. Minahan, ``{Circular semiclassical string solutions on AdS(5) x S(5)},''
  {\em Nucl.Phys.} {\bf B648} (2003) 203--214,
  \href{http://xxx.lanl.gov/abs/hep-th/0209047}{{\tt hep-th/0209047}}.

\bibitem{website}
\url{http://www.thphys.uni-heidelberg.de/segmented-AdS-strings}.

\bibitem{vRegSquare}
\url{http://www.thphys.uni-heidelberg.de/~holography/segmented-strings/assets/videos/reg_square/video.html}.

\bibitem{vRegHex}
\url{http://www.thphys.uni-heidelberg.de/~holography/segmented-strings/assets/videos/reg_hex/video.html}.

\bibitem{vRegOct}
\url{http://www.thphys.uni-heidelberg.de/~holography/segmented-strings/assets/videos/reg_oct/video.html}.

\bibitem{vIrregHex}
\url{http://www.thphys.uni-heidelberg.de/~holography/segmented-strings/assets/videos/irreg_hex/video.html}.

\bibitem{Vegh:2015ska}
D.~Vegh, ``{The broken string in anti-de Sitter space},''
  \href{http://xxx.lanl.gov/abs/1508.06637}{{\tt 1508.06637}}.

\end{thebibliography}\endgroup
\end{document}